\documentclass[12pt]{article}

\usepackage{amsmath}
\usepackage{epsf}
\usepackage{graphicx}

\begin{document}
\sloppypar

\date {}
 {\it Accepted for publication in Astronomy
Letters, v 26, p 699, 2000.}
\vspace{2cm}
\bigskip
 
\large
\centerline{\bf Energy Release During Disk Accretion onto a Rapidly Rotating Neutron 
Star}
\vspace{15mm}
 \normalsize
\centerline{ Nail Sibgatullin$^{1,2}$ and Rashid A. Sunyaev$^{2,3}$}
\vspace{2mm}
\noindent
$^1${\it Moscow State University, Vorob'evy gory, Moscow, 119899 Russia }\\
$^2$ {\it Max-Planck-Institut f\"ur Astrophysik,
Karl-Schwarzschild-Str. 1, 85740 Garching bei \\
 \noindent Munchen, Germany}\\
$^3$ {\it Space Research Institute, Russian Academy of Sciences, ul. 
Profsoyuznaya 84/32, Moscow, 117810 Russia } \\
$^*$ {e-mail:sibgat@mech.math.msu.su}
 
\vspace{7mm}
\def\beq{\begin{equation}}
\def\f{\frac}
\def\p{\partial}
\def\eeq{\end{equation}}
\noindent
{\small ...Abstract }
  The energy release $L_s$ on the surface of a neutron star 
(NS) with a weak magnetic field  and the energy release $L_d$ in the 
surrounding accretion disk depend on two independent parameters that 
determine its state (for example, mass $M$ and cyclic rotation frequency 
$f$) and is proportional to the accretion rate. We derive simple 
approximation formulas illustrating the dependence of the efficiency of 
energy release in an extended disk and in a boundary layer near the NS 
surface on the frequency and sense of rotation for various NS equations 
of state. Such formulas are obtained for the quadrupole moment of a NS, 
for a gap between its surface and a marginally stable orbit, for the 
rotation frequency in an equatorial Keplerian orbit and in the 
marginally stable circular orbit, and for the rate of NS spinup via disk 
accretion. In the case of NS and disk counterrotation, the energy 
release during accretion can reach $ 0.67\dot{M}c^2.$ The sense of NS rotation is 
a factor that strongly affects the observed ratio of nuclear energy 
release during bursts to gravitational energy release between bursts in 
X-ray bursters. The possible existence of binary systems with NS and 
disk counterrotation in the Galaxy is discussed. Based on the static 
criterion for stability, we present a method of constructing the 
dependence of gravitational mass $M$ on Kerr rotation parameter $j$ and on 
total baryon mass (rest mass) $m$ for a rigidly rotating neutron star. We 
show that all global NS characteristics can be expressed in terms of the 
function $M(j, m)$ and its derivatives. We determine parameters of the 
equatorial circular orbit and the marginally stable orbit by using $M(j, 
m)$ and an exact solution of the Einstein equations in a vacuum, which 
includes the following three parameters: gravitational mass $M$, angular 
momentum $J$, and quadrupole moment $\Phi_2.$  Depending on $\Phi_2$, this 
solution can also be interpreted as a solution that describes the field 
of either two Kerr black holes or two Kerr disks.


 \pagebreak

\section*{Introduction}

{\bf Observational Facts}. Three independent observational facts have 
prompted us to revert to the problem of disk accretion onto neutron 
stars (NSs) with weak magnetic fields, which have virtually no effect on 
the accretion dynamics.

(1) The discovery of an accreting X-ray pulsar/burster with rotation 
period $p = 2.8 \mbox{ms} $ (cyclic frequency $f = 1/p = \Omega/2\pi 
= 401 Hz$) in a binary system with an orbital period of 2 h, SAX J1808.4 - 3658 (Van der Klis et al. 2000; Chakrabarty and Morgan 1998; 
Gilfanov et al. 1998).

(2) The detection of quasi-periodic oscillations in X-ray bursters 
during X-ray bursts with frequencies of 300 - 600 Hz from the RXTE 
satellite (Stromayer et al., 1998; Van der Klis et al., 2000). In the 
pattern of slow (in seconds!) motion of the nuclear helium burning front 
over the stellar surface, these flux oscillations can be naturally 
interpreted as evidence of rapid NS rotation with the oscillation 
frequency of the X - ray flux. The rotation frequencies of the X-ray 
bursters KS 1731 - 260, Aql X-1, and 4U 1636 - 53 are 523.9, 548.9, and 
581.8 Hz, respectively (Van der Klis et al., 2000). The rotation 
periods of these neutron stars are close to 1.607 ms, the rotation 
period of the millisecond radio pulsar B1957+20, the shortest one among 
those found to date (Thorsett and Chacrabarty 1998). Recall that 
B1957+20 is a member of a binary system with a 0.362 day period.

(3) The discovery of twenty millisecond radio pulsars in the globular 
cluster 47 Tuc. Most of these pulsars have periods from 2 to 8 ms and 
are members of close low-mass binaries. The total number of millisecond 
pulsars in 47 Tuc is estimated to be several hundred (Camilo et al., 
1999).

Measurements of the spindown rate for millisecond pulsars attest to 
magnetic fields of $10^8 - 10^9 G.$ Neutron stars with appreciable magnetic 
fields may manifest themselves as millisecond radio pulsars; NSs with 
weaker fields are simply unobservable at the current sensitivity level 
of radio telescopes, but nothing forbids their existence. 

Disk accretion onto a NS with a weak magnetic field $ H < 10^8 G$, 
proceeds in 80 low-mass binaries of our Galaxy. Strong fields could 
affect the accretion dynamics and could give rise to periodic X-ray 
pulsations, which is not observed in these systems.

{\bf Accretion Pattern}. Accreting matter with a large angular momentum 
spins up a neutron star (Pringle and Rees 1972; Bisnovatyi-Kogan and Komberg 1974; Alpar et al., 1982; Lipunov and Postnov, 1984) and causes 
appreciable energy release in the accretion disk in a boundary layer 
near the NS surface (Shakura and Sunyaev, 1988; Popham and Sunyaev, 2000). 
Inogamov and Sunyaev (1999) considered the formation of a layer of 
accreting matter spreading over the surface of a neutron star without a 
magnetic field. The surface radiation was found to concentrate toward 
two bright rings equidistant from the stellar equator and the disk 
plane. The distance of the bright rings from the equator depends on the 
accretion rate alone.

In the course of accretion, the baryon and gravitational masses of the 
NS grow, its rotation velocity and moment of inertia change, a 
quadrupole component appears in its mass distribution, and its external 
gravitational field changes. For several standard equations of state of 
matter, the NS equatorial radius proves to be smaller than the 
equatorial radius of a marginally stable orbit over a wide range of 
rotation frequencies. This significantly affects the ratio of energy 
release in an extended disk and the stellar surface. The X-ray spectra 
of the accretion disk and the spread layer can differ greatly, which 
opens up a possibility for experimentally testing the theoretical 
results presented below.

{\bf Energy Release in the Disk and on the NS Surface.}  Here, we calculate 
the total energy release during accretion onto a rapidly rotating NS and 
determine the ratio of disk luminosity $L_d$ to luminosity $L_s$ of the 
spread layer on the stellar surface at a given accretion rate $M.$ 
Multiplying the efficiency of energy release $\epsilon_{grav}$ by $\dot{M}$ yields the sought-for luminosities. Clearly, allowance for radiation-pressure 
forces and for the detailed boundary-layer physics can slightly modify 
the derived formulas (Marcovic and Lamb, 2000; Popham and Sunyaev, 2000).

For the most important case of a NS with fixed gravitational mass $M =
1.4 M_{\odot}$, we present the results of our calculations in Fig. 1. The 
calculations were performed for a moderately hard equation of state (EOS 
FPS). Below, we use the notation of Arnett and Bowers (1977) for EOS A 
(Pandharipande, 1981), L, and M; EOS AU from Wiringa et al. (1988); and 
EOS FPS from Lorenz et al. (1993). EOS FPS is a modern version of the 
equation of state proposed by Friedman and Pandharipande (1981).

The approximation formula (derived in section 5) for the total NS 
luminosity as a function of cyclic rotation frequency (see Fig. 1) is
\begin{equation}
L_s + L_d\approx (0.213 - 0.153\,( f/1\,\mbox{kHz}) + 0.02\, (
f/1\,\mbox{kHz})^2) \dot{M} c^2.
\end{equation}
 
Here, $f$ varies in the range from -1 to +1 kHz, with positive and 
negative $f$ corresponding to NS and accretion-disk corotation and 
counterrotation, respectively. In this paper, a large proportion of the 
results in graphical form and in the form of approximation formulas are 
given for a NS gravitational mass of 1.4 $M_{\odot}$ or for normal sequences with a 
 rest mass whose gravitational mass is 1.4 $M_{\odot}$ in the static limit 
for various equations of state of the matter in the NS interior. 
Amazingly, the measured mass $M$ of an absolute majority of the 
millisecond pulsars in binaries lies, with high reliability, in a narrow 
range $1.35 \pm 0.04 M_{\odot}$ (Thorsett and Chakrabarty 1998). Note, however, 
\begin{figure}[t]
\includegraphics[width=0.75\linewidth]{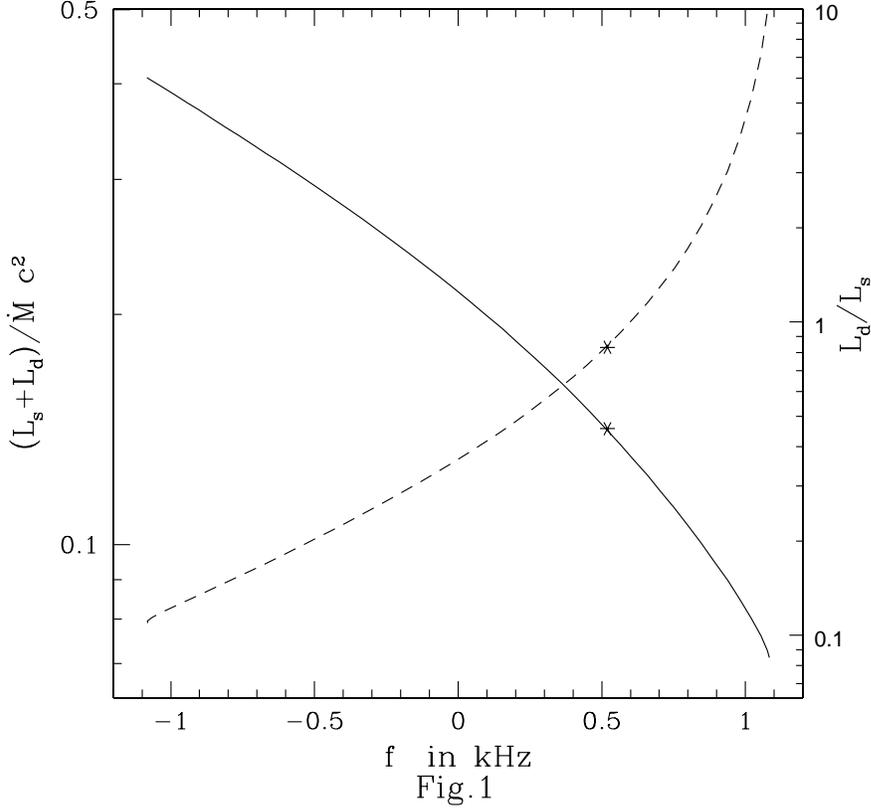}
\caption{\small  Total energy release  $L_s + L_d$  on the NS surface and in the 
accretion disk (solid line) and ratio of energy releases in the disk $L_d$ 
and on the surface $L_s$ (dashed line) versus NS rotation frequency $f$ at 
the fixed gravitational mass $ M = 1.4 M_{\odot}$  for the moderately hard EOS 
FPS. The asterisks in Figs. 1 - 4 correspond to the frequency $f$ at which 
the NS radius is equal to the radius of the marginally stable orbit. At 
$f\leq f_*$, there is a gap between the accretion disk and the NS surface 
[see formula (3)]. }
\end{figure}
that the mass of the X-ray pulsar VELA X-1 is close to 1.8 $M_{\odot}$.

When observational data are interpreted, it is useful to have an 
approximation formula (EOS FPS) for the ratio of the NS surface and 
total luminosities,
\begin{equation}
L_s/(L_s + L_d) \approx 0.737 - 0.312\,( f/1\,\mbox{kHz}) - 0.19\,
(f/1\,\mbox{kHz})^2. 
\end{equation}

We define the efficiency of energy release in the disk as the binding 
energy of a particle in a Keplerian orbit at the inner disk boundary. 
This orbit coincides with the marginally stable orbit or with the orbit 
at the NS equator.

For a NS of mass $M = 1.8 M_{\odot}$ with EOS FPS, the calculated total luminosity and $L_d/L_s$ are shown in Fig. 2. The corresponding 
approximation formulas are given in section 5.
\begin{figure}[t]
\includegraphics[width=0.75\linewidth]{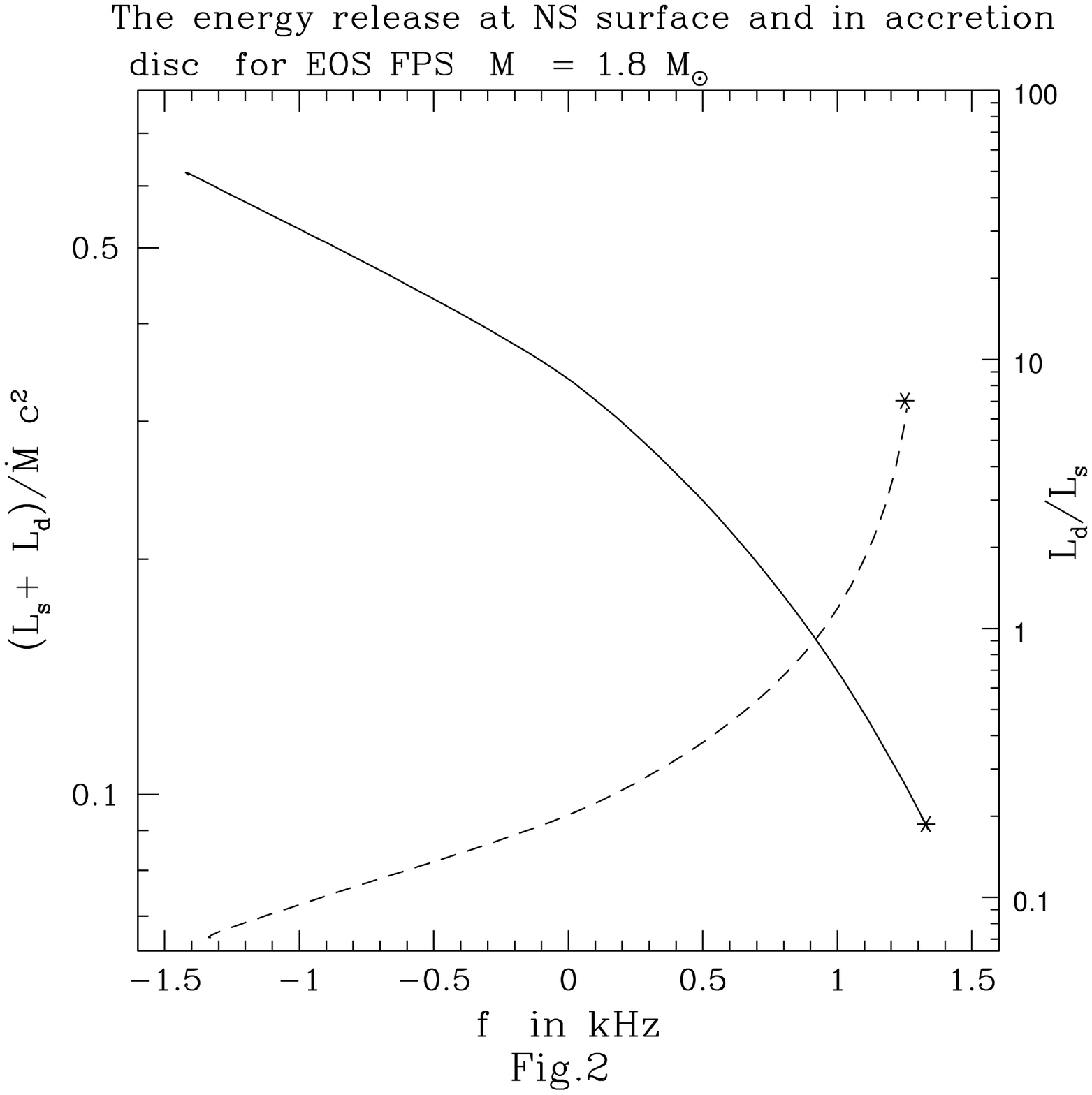}
\caption{\small Total energy release on the NS surface and in the accretion disk 
(solid line) and ratio of the disk and surface energy releases (dashed 
line) versus NS rotation frequency for the fixed gravitational mass $ M 
= 1.8 M_{\odot}$ (EOS FPS). The NS is stable to gravitational collapse only in 
the presence of rotation. }
\end{figure}
We see from Figs. 1 and 2 that the surface luminosity $L_s$ dominates over 
the disk luminosity for a slowly rotating star and in the case of 
counterrotation. Note that the effective energy release during accretion 
onto a NS of gravitational mass $M = 1.8 M_{\odot}$ reaches 0.62 $ Mc^2$ (at a 
cyclic frequency of NS rotation in the sense opposite to disk rotation 
equal to 1.42 kHz). For a normal sequence with a maximum mass losing 
stability in the static limit, the total energy release even reaches 
$0.67\dot{M}c^2$ (see Fig. 3). Note that the above values exceed appreciably 
the disk energy release $E_d = 0.422 \dot{M}c^2$ during accretion onto a Kerr 
black hole with the largest possible rotation parameter $j = 1.$ 
Clearly, such a high energy release is also associated with the loss of 
kinetic energy of stellar rotation during accretion of matter with an 
oppositely directed angular momentum.
\begin{figure}[t]
\includegraphics[width=0.75\linewidth]{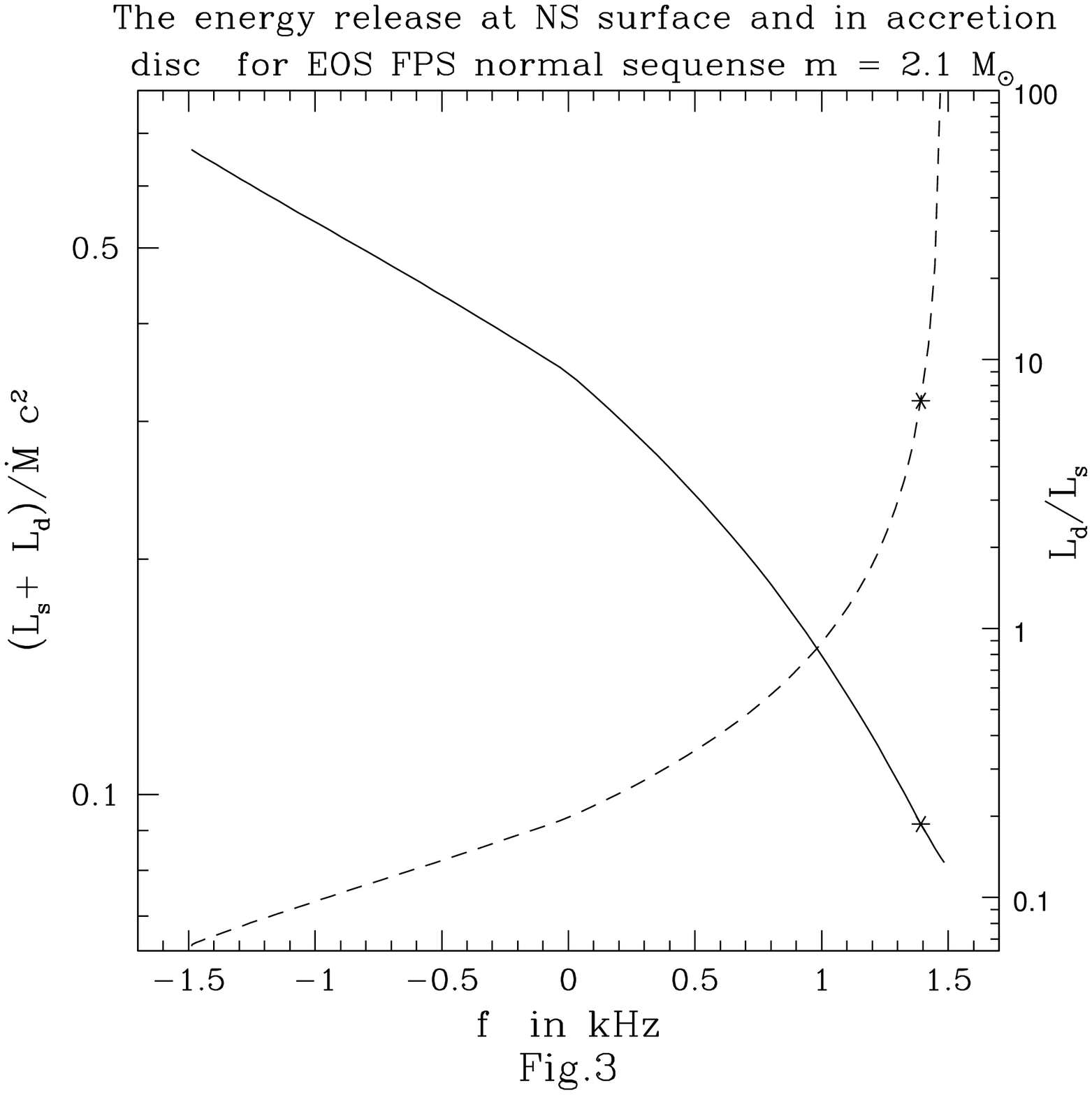}
\caption{\small  Total energy release on the NS surface and in the accretion disk 
y(solid line) and ratio of the disk and surface energy releases (dashed 
line) versus NS rotation frequency for the EOS FPS normal sequence with 
 $m = 2.1 M_{\odot}$ or 1.8 $ M_{\odot}$ in the static limit. }
\end{figure}
The accretion-disk luminosity at $f > 600 Hz$ exceeds the surface 
luminosity if the star rapidly rotates in the same sense as does the 
disk. We see from Fig. 1 that $L_s \approx L_d \approx 0.064 Mc^2$ at $f = 600 Hz.$

This tendency is seen from simple Newtonian formulas (Kluzniak
1987; Kley 1991; Popham and Narayan 1995; 
Sibgatullin and Sunyaev 1998, below referred to as SS 98):
$$
L_s = \f{1}{2}\dot{M} G M (1- f/ f_K)^2/R,\quad L_d =\f{1}{2}\dot{M} G M/ R.
$$
Here, $R$ is the equatorial stellar radius, and $f_K =\sqrt{ G M/R^3}/(2\pi)$ is the 
Keplerian rotation frequency at the inner disk boundary. This formula is 
valid at low angular velocities, when, to a first approximation, NS 
oblateness can be disregarded. The exact formulas for $L_s$ and $L_d$ when 
 the disk corotate lies in
equatorial plane of star are (Sibgatullin and Sunyaev 2000; below 
referred to as SS 00)
\begin{equation}
L_s = 2\pi^2\dot{M}  R^2 (f_K - f)^2,\quad L_d =\dot{M}\f{1}{2R}
\f{d \phi(r) r^2}{d r}|_{r=R}\tag{$1a$}
\end{equation}

Here, $\phi(r)$  is the gravitational potential in the disk plane as a 
function of distance from the NS. In particular, it follows from formula 
(1a) that, at $|f| = 0.5 f_K$, the surface energy release for 
counterrotation is a factor of 9 greater than that for NS and disk corotation ! 
\begin{figure}[t]
\includegraphics[width=0.75\linewidth]{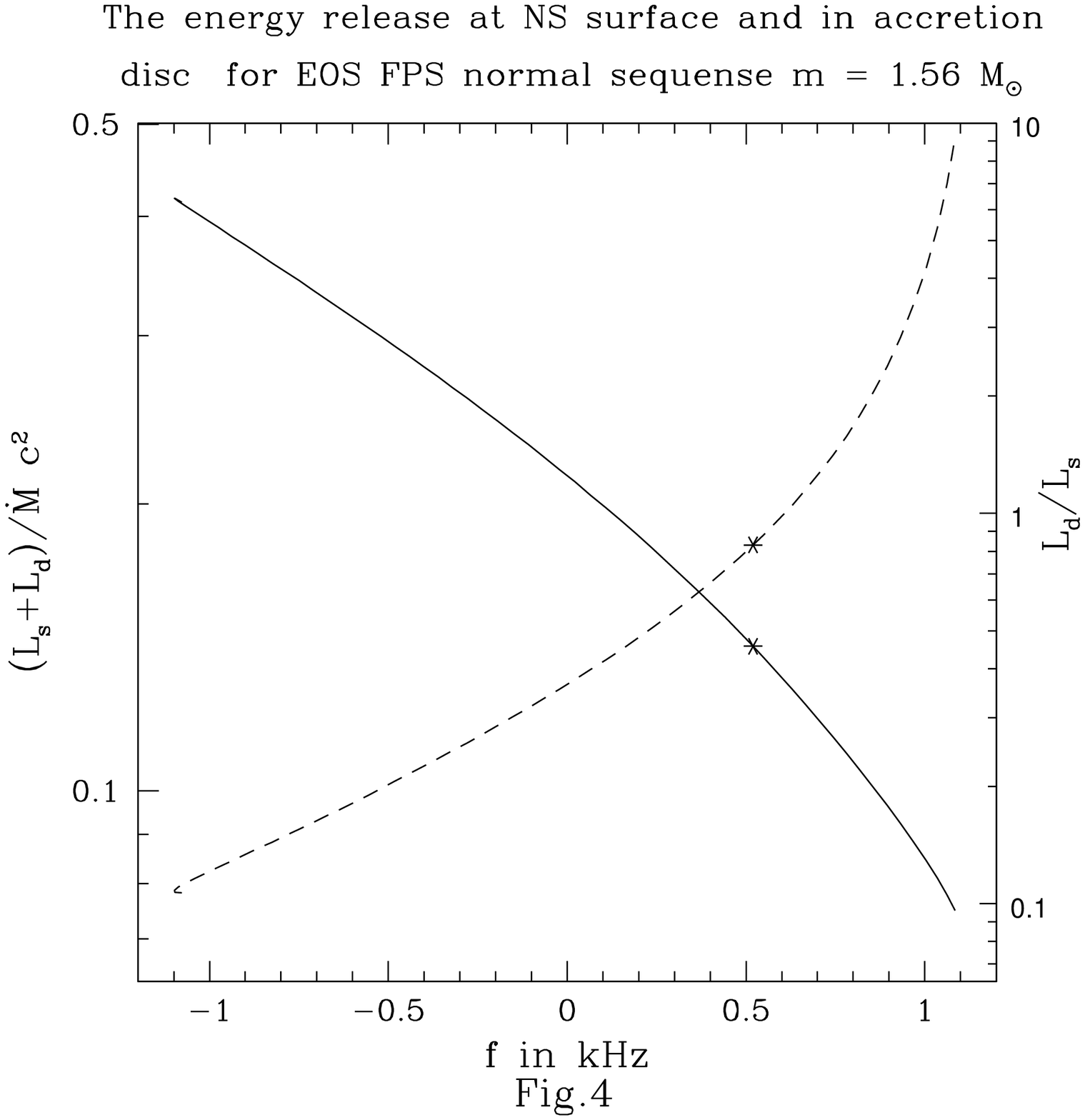}
\caption{\small  Total energy release on the NS surface and in the accretion disk
(solid line) and ratio of the disk and surface energy releases (dashed line)
versus NS rotation frequency for the EOS FPS normal sequence with $m = 1.56
M_{\odot} (1.4 M_{\odot}$ in the static limit). The asterisks correspond to
the rotation frequency at which $R = R_*.$}
\end{figure}
In Fig. 4, we present similar results of our calculations for stars of 
fixed rest mass (total baryon mass) $m$ corresponding to $M = 1.4 M_{\odot}$ in 
the static limit for the NS EOS FPS. The difference between this figure 
and Fig. 1 is not large, because the increase in gravitational mass 
through rapid rotation is relatively small [but appreciable; see 
formulas (18)].

For the fixed gravitational mass $M = 1.4 M_{\odot}$, we derived a simple 
approximation formula for the gap between the equator of a rotating star 
with radius $R$ and a marginally stable orbit for EOS FPS with radius 
$R_*$:
\begin{equation}
(R_* -R)/1\mbox{km}\approx 1.44 -3.061\,( f/1\,\mbox{kHz})
+ 0.843\,(f/1\,\mbox{ËçÃ})^2 + 0.6\,(f/1\,\mbox{kHz})^3
 - 0.22\,(f/1\,\mbox{kHz})^4 
\end{equation}

 {\bf Using the Static Criterion for Stability.} Here, we attempt to derive an apprixmation formula for the dependence of NS gravitational mass 
$M(j, m)$ on Kerr rotation parameter $j \equiv cJ/GM^2$ (where $J$ is the NS 
angular momentum) and on its rest mass $m.$

We derive the function $M(j, m)$ (which is given below for two NS 
equations of state) by using the static criterion for the loss of 
stability and data obtained using the numerical code of Stergioulas 
(1998). Knowledge of $M(j, m)$ allowed us to derive formulas for the NS 
angular velocity and equatorial radius as functions of $j$ and $m.$

{\bf Metric Properties of a Rotating Neutron Star.} Remarkably, the external 
field of a rapidly rotating NS with a mass larger than the solar one can be satisfactorily described in terms of general relativity by 
introducing only one additional parameter compared to the Kerr metric 
- quadrupole moment of the mass distribution (SS 98). In sections 2 and 3, we discuss exact solutions that take 
into account higher multipole moments for low masses.

Analogous to $M(j, m)$, we managed to construct approximation dependences 
of the additional (to the Kerr one) dimensionless quadrupole coefficient $b$ for the external gravitational field on $j$ and $m.$ Using these 
approximations in an exact solution for the metric outside rigidly 
rotating NSs enabled us to analytically calculate parameters of the 
marginally stable orbit in the accretion disk, energy release in the 
disk and on the NS surface, and the rate of NS spinup via disk 
accretion. In order to relate the derived approximation dependences to 
the observed NS parameters, we passed from the Kerr parameter $j$ to the 
observed parameter $f$ (cyclic rotation frequency) for 1.4 $M_{\odot}$ stars in the 
final formulas. In section 1, we give approximation formulas for the 
relationship of $j$ to $f$ and $M.$

{\bf The Content of the Paper.} Below (in section 1), we present a method 
for global construction of the NS gravitational mass as a function of 
its Kerr rotation parameter $j$ and rest mass $m$. The static criterion 
for stability underlies the method. The results of application of our method
 approximate the numerical
data obtained with help of the numerical code of Stergioulas (1998)  very accurately.

The NS angular velocity and its equatorial radius are determined by 
using $M(j, m).$ We propose a new formula for the equatorial radius, 
which matches the exact one at $j = 0.$

In section 2, we present a method of constructing the quadrupole 
coefficient via $M(j, m)$ using an exact solution of the Einstein 
equations for the metric of the external gravitational field.

The exact solutions that describe the external fields of rigidly 
rotating stars with arbitrary multipole structure are discussed in 
section 3.

Global properties of the exact quadrupole solution are described in 
section 4. We note that the corresponding gravitational field in some 
region on the $b, j$ plane  
(including low values of the quadrupole moment $b$ and the Kerr rotation 
parameter $j$) behaves as the field of two rotating black holes, with the NS 
pressure acting as an elastic support. However, outside this region, the 
solution properties outside the star are equivalent to the field of two 
supercritical Kerr disks. By contrast to black holes (Hoenselaers 1984), 
two Kerr disks can be in equilibrium in the absence of supports (Zaripov 
et al. 1994; a graphic post-Newtonian approach was developed by 
Zaripov et al. 1995).

In section 4, we derive expressions for the energy, angular momentum, 
radius, and angular velocity of particles in the marginally stable orbit 
in the quadrupole solution. The above functions depend on $j$ and $b.$ At 
$b = 0$ and $j = 1$, these expressions have numerical solutions which 
were first found by Ruffini and Wheeler (1970). At $b = 0$ and $j < 1,$ 
the above expressions approximate the corresponding formulas of Bardeen 
et al. (1972) in the form of polynomials in $j.$

The energy, angular momentum, radius, and Keplerian angular velocity of 
particles at the stellar equator depend markedly on the NS equation of 
state. In section 4, these functions are given as functions of $j, m $
(and $j, M$) for EOS FPS and as functions of $j, m$ for EOS A.

The above set of functions proves to be enough to calculate the 
dependences of energy release on the NS surface and in the accretion 
disk (section 5) on $j, m$ or on $j, M$. In section 5, we provide 
approximation formulas for the total luminosity and the luminosity ratio 
of the disk and the NS surface as functions of rotation frequency at 
fixed gravitational masses $M = 1.4$ and $1.8 M_{\odot}$ for EOS FPS. Similar 
approximation formulas are given for the EOS A and EOS AU normal 
sequences with $M = 1.4 M_{\odot}$ in the static limit. In section 5, we also 
discuss the NS spinup and provide approximation formulas for the spinup 
rate in the case of EOS FPS for fixed gravitational masses of 1.4 and 
$1.8 M_{\odot}.$ For comparison, we give formulas for the dependence of 
luminosity and spinup parameter on the dimensionless rotation frequency 
in the Newtonian theory.

Astrophysical applications and implications of our results (particularly 
for the most interesting case of NS and disk counterrotation) are 
discussed in section 6.

\section{The Static Criterium for Stability and the Function $M(j,m)$.}
The static stability criterion for nonrotating isentropic stars 
(planets) has been discussed in the literature since the early 1950s 
(Ramsey 1950; Lighthill 1950; Zel'dovich 1963; Dmitriev and Kholin 1963; 
Calamai 1970). It was generalized to rotating configurations by
Bisnovatyi-Kogan and Blinnikov (1974) and
Hartle (1975).

Below, we follow Zel'dovich (1963) in its interpretation. In what 
follows, $M$ is the NS gravitational mass, and $m$ is the total mass of
its constituent baryons (rest mass). All masses are measured in solar 
masses, so equalities of the type $m = 1.16$ and $M = 1.4$ imply that 
$m = 1.16 M_{\odot}$ and $M = 1.4 M_{\odot}$, respectively.

We choose central density $\rho$ as one of the independent arguments and 
angular momentum $J$ as the second argument. It can then be shown that 
the extremum of gravitational mass $M$ in central density at fixed angular momentum coincides with the extremum of rest mass $m$ (Zel'dovich 
and Novikov 1971; Shapiro and Teukolsky 1985). At the extremum, a 
steady-state configuration becomes unstable to the neutral mode of 
quasi-radial oscillations. NS stable states can exist only at lower 
masses.

Denote the extremum central density at fixed angular momentum by $\rho_* =
\rho_*(J)$. The functions  $M(\rho,J)$  and $ m(\rho, J)$  near the extremum at fixed 
angular momentum can then be expanded in Taylor series:
\begin{equation} 
 M(\rho, J) = M_* + a_1(\rho -\rho_*)^2 + a_2(\rho - \rho_*)^3 +a_3(\rho - \rho_*)^4
+\dots,  
\end{equation}
\begin{equation} m(\rho, J) =m_* + b_1(\rho - \rho_*)^2 + b_2(\rho - \rho_*)^3 + b_3(\rho - \rho_*)^4
+\dots
\end{equation}
The coefficients $\,a_i,\,b_i, \quad i = 1,2 \dots$   in formulas (4) and (5) depend on $J.$

Equation (5) allows the central density $\rho$ to be expressed as a 
power series of $\sqrt{ m_*-m} $:
$$\rho =\rho_* + \rho_0(m_*-m)^{0.5} + \rho_1(m_* - m) + \rho_2(m_*-m)^{1.5}
 +\dots
$$ 
Substituting this expression for $\rho$ in (4) yields
\begin{equation}
M -M_* = M_0(m_*-m) + M_1(m_* - m)^{1.5} + M_2(m_*-m)^2
 +\dots
\end{equation}
In this formula, $ M_*,\quad m_*,\quad M_0,\quad M_1,\quad M_2,\dots $ are some functions of 
angular momentum $J.$

If, alternatively, the central density is expressed as a series in 
half-integer powers of $ M -M_*$  using (4) and substituted in (5), then
\begin{equation}
m -m_* = m_0(M_*-M) + m_1(M_* - M)^{1.5} + m_2(M_*-M)^2
 +\dots
\end{equation}
Note the formal similarity between the expansions (6), (7) and the 
expansions for the mass and radius near the point of phase transition at 
the stellar center in nonrotating stellar models with phase transitions 
in general relativity (Seidov 1971; Lindblom 1998).

{\bf Determining the Gravitational and Rest Masses at the Stability 
Boundary.} The numerical code of Stergioulas (1998) allows NS parameters to be computed for a given equation of state by specifying a numerical 
value of the central density and one of the parameters $M,\,m,\,\Omega, J,\,\Omega_K$. The functions $M_*(J)$ and $m_*(J)$ can be determined as 
the maximum values of $ M(\rho, J), m(\rho, J)$  for any fixed $J$ by using this 
numerical code.

It is convenient to use a dimensionless angular momentum $j \equiv cJ/GM^2$ 
(Kerr rotation parameter) instead of the angular momentum $J,$ because, 
apart from the mass, the metric of the external field of a rigidly 
rotating neutron star is determined precisely by this parameter. The 
functions $ M_*(j),\quad m_*(j)$ have the meaning of dependences of the 
gravitational and rest masses on Kerr parameter at the stability 
boundary.

We approximate $M ( j, m),\quad m(j,M)$  by the above terms of the expansions (6) 
and (7) in finite ranges of the parameters: $ |j|\le 0.7,\quad 1.1
M_{\odot} <m< 2.5M_{\odot}$. For this approximation, the extremum property of the 
gravitational and rest masses  at loss of stability  is retained . 
It is this property that is the main idea behind the static criterion 
for stability.

{\bf Determining the NS Angular Velocity and its Moment of Inertia.} In SS 
00, we showed that
 \begin{equation}
\Omega =2 \pi f=\f{ \p
  M c^2}{\p
J}|_{m},
\end{equation}
where $|_m$ denotes, as usual, differentiation at constant $m.$ Given the 
definition of $j,$  we have from (8)
$$ 
\Omega =2\pi f=\f{ \p M c^2}{\p
J}|_{m}=  \f{\p M}{\p j}|_{m} c^3/GM^2 - 2 j\f{ \p M c^2}{\p
J}|_{m}/M,
$$
hence
\begin{equation}
\Omega=\f{ c^3}{G M}  M_{,j}|_{m}/\left(M +2 j  M_{,j}|_{m}\right);\qquad
M_{,j}|_{m}\equiv  \f{\p M}{\p j}|_{m}.
\end{equation}
In what follows, a comma in the subscript denotes a partial derivative 
with respect to the corresponding argument. Using $M(j, m)$, we can also 
determine the moment of inertia
$$
I = \left(\Omega_{, J}\mid_m)\right)^{-1} =
c^2\left( M_{JJ}|_{m}\right)^{-1}
=\f{G}{c}(M^2 +2 j M M_{,j}\mid_m)/\Omega_{, j}\mid_m .
$$
At the stability boundary  $m = m_*(j)$, we have from (6), (9)
\begin{equation}
\Omega_*(j)=\f{ c^3}{G M_*(j)
}\f{(M_*(j))^{\prime} + M_0(j)(m_*(j))^{\prime}}{(M_*(j) +2 j (M_*(j))^{\prime} + M_0(j)(m_*(j))^{\prime} )}.
\end{equation}
Here, the prime denotes a derivative of the corresponding function with 
respect to $j.$

{\bf Determining the Coefficients $M_0(j), M_1(j), M_2(j)$.}  The 
coefficients $M_0(j), M_1(j), M_2(j)$ in formula (6) are sought in the 
form of formal expansions in even powers of $j$. We calculate the 
limiting values of these function when $j\rightarrow 0$   by approximating the 
dependence of gravitational mass on rest mass in the static case:
\begin{equation}
M_{st} =M_*(0)+M_0(0) (m_*(0) - m) + M_1(0)(m_*(0) - m)^{1.5} + M_2(0)(m_*(0) - m)^2\dots
\end{equation}
Based on numerical data for $\Omega$  at the stability boundary  $m = m_*(j)$, we determine $M_0(j)$ from formula (10) by using the derived 
functions  $ M_*(j),\, m_*(j).$

In order to successively determine the remaining coefficients
$M_1(j),\,M_2(j)$ before the nonstatic terms in (6), we introduce the coefficient $\mu_1 - m$ , with $M_1(j)$ (whose value at $j = 0$ is already known) being 
calculated by using numerical data precisely at $ m =\mu_1$.  We then 
have
\begin{equation}
M_1(j)=\f{ M(j,\mu_1) -
M_0(j)(m_*(j)-\mu_1)-M_2(0)(m_*(j)-\mu_1)^2}{(m_*(j)-\mu_1)^{1.5}}.
\end{equation}
Having numerical data for the right-hand part at discrete $j$, we can 
easily find a sixth-degree polynomial with the smallest rms deviation by 
points. In our calculations, we choose  $\mu_1$  in such a way that $ M_{st}(\mu_1)$ is equal to or differs only slightly from 1.4 $M_{\odot}.$ 

Having derived the expression for $M_1(j)$, we can determine $M_2(j)$ by 
using a different normal sequence, say, at  $m = \mu_2.$

We construct the coefficient $M_2(j)$  as follows:
\begin{eqnarray}
M_2(j)=\f{ M(j,\mu_2) -
M_0(j)(m_*(j)-\mu_2)-M_1(j)(m_*(j)-\mu_2)^{1.5}}{(m_*(j)-\mu_2)^2}.
\end{eqnarray}

{\bf The Function $M(j, m)$ for EOS A.} We choose the following constants 
 $\mu_1,\mu_2 :\quad \mu_1=1.5663, \mu_2=1.1663$ for EOS A. The 
numerical data of Stergioulas's code for the dependences of critical 
masses on $j$ can be approximated as
$$
M_*(j) = 1.659 + 0.489\, j^2 + 
    0.273\, j^4 + 0.32\, j^6,
$$
\begin{equation}
m_*(j) =  1.9495  +  0.5144\, j^2 + 
    0.335\, j^4 + 
    0.38\, j^6.
\end{equation}
Combining our results (10 - 14) for EOS A, we finally obtain
\begin{eqnarray}
M(j, m) \approx & M_*(j) +(-0.6473  - 0.019\,j^2 - 
 0.182\, j^4 + 0.23\, j^6)\,( m_*(j) - m) + \nonumber
\\&(-0.0808 - 0.039\, j^2 + 0.31\, j^4 - 0.34\, j^6 )\,( m_*(j)
 - m)^{1.5}+\nonumber
\\&(-0.0457 + (0.059\, j^2  -0.222\, j^4 +0.28\, j^6 )\,(1.5663 - m))\,( m_*(j) - m)^2.
\end{eqnarray} 
Formula (15) describes the data of Stergioulas's (1998) numerical code 
to within the fourth decimal place in mass (expressed in solar masses) and to within the third decimal place in angular velocity $\Omega$
calculated using (9) (and expressed in units of $10^4 $ rad/ s) in the 
following parameter ranges:  $\quad
-0.7< j < 0.7,\quad 1.1 < m < 2.4.$ 

{\bf The Function $M(j, m)$ for EOS FPS}. For EOS FPS, which is stiffer than 
EOS A, our numerical searches for the maximum gravitational and rest 
masses at fixed angular momentum lead to the following dependences of 
these masses on rotation parameter at the stability boundary:
$$
 M_*(j) =1.8016+ 0.572\, j^2 + 
    0.164\, j^4 + 0.54\, j^6 ;$$
\begin{equation}
 m_*(j)=2.104 + 0.619\, j^2 + 0.154\, j^4 + 
    0.71\, j^6.
\end{equation}
Approximating the data for the static case $j = 0$ yields
$$
M_{st} = 1.8016 - 0.6546\,(2.104 - m)  - 0.0846\,(2.104 -
 m)^{1.5} - 0.035\, (2.104 - m)^2 \dots
$$
We determine $ M_0(j),\, M_1(j),\, M_2(j)$  by the above procedure: for the 
rest masses of normal sequences, we choose  $\mu_1=1.56, \quad \mu_2=1.16 $. 
Of course, the choice of these values is rather arbitrary. The final 
formula derived from (10 - 13) and (16) for the dependence of 
gravitational mass on rotation parameter and rest mass for EOS FPS is
\begin{eqnarray}
M(j, m ) \approx & M_*(j)+ (-0.6538 - 0.0295\, j^2 - 
 0.242\, j^4 + 0.49\, j^6)\,( m_*(j) - m) +\nonumber
 \\&(-0.0846 - 0.009\, j^2  + 
 0.302\, j^4  - 0.52\, j^6 )\,( m_*(j) - m)^{1.5} +\nonumber
 \\&(-0.035 + ( 0.039\, j^2  - 0.188\, j^4  + 0.32\,
 j^6)(1.56 - m ))\,( m_*(j) - m)^2.
\end{eqnarray}
Formula (17) for the gravitational mass, like the previous (15), approximates the numerical data of Cook et al. (1994) and the results
of Stergioulas's (1998) code in the argument ranges $-0.7 < j < 0.7,\quad
1.1< m <2.5$ with an amazing accuracy: to within the fourth decimal 
place in gravitational mass and to within the third decimal place in 
angular velocity (expressed in units of $10^4$ rad/ s).

{\bf The Function $m(j, M)$ for EOS FPS.} Similarly, we can approximate $m(j, M)$ by using numerical data for the angular velocity, the baryon and 
gravitational masses at the stability boundary, and for the fixed $M = 1.4 M_{\odot}$ and $ M =
M_{\odot}$. For EOS FPS, we then obtain
\begin{eqnarray*}
 m(j, M)\approx & m_*(j) +  (-1.528 + 0.059\, j^2 + 0.45\, j^4 - 0.72\, j^6)\,( M_*(j) - M) +\nonumber
 \\& 
 (0.269 - 0.076\, j^2 - 0.354\, j^4 + 0.6\,j^6)\,
   ( M_*(j) - M)^{1.5} +\nonumber
 \\& 
 (0.004 +  (-0.121\, j^2 + 0.68\, j^4 - 1.05\, j^6)\,(1.4 - M))
 ( M_*(j) - M)^2.
\end{eqnarray*}

{\bf The Dependence of Gravitational Mass and Dimensionless Angular Momentum 
$j$ on Angular Velocity for 1.4 $M_{\odot}$  Normal Sequences in the Static Limit.} 
At present, the gravitational masses and angular velocities of neutron 
stars are measured with a high accuracy (Thorsett and Chakrabarty 1998). 
Here, we give, for reference, approximations of $M(f)$ and $j(f)$   for 
various equations of state for $ M =1.4 M_{\odot}$ normal sequences in the 
static limit (here, the masses are in solar masses, and  $f=\Omega/2\pi $) is 
the cyclic rotation frequency):
$$\mbox{EOS A}:\qquad
M \approx 1.4 +0.0075 \,(f/1\mbox{kHz})^2 +0.0055 \,(f/1\mbox{kHz})^4,
$$
$$
j \approx 0.274 \,(f/1\mbox{kHz})+0.14 \,(f/1\mbox{kHz})^2,\qquad m=1.566;
$$
$$
\mbox{EOS AU}:\qquad
M \approx 1.4 +0.01 \,(f/1\mbox{kHz})^2 +0.006 \,(f/1\mbox{kHz})^4,
$$
$$
j \approx 0.34 \,(f/1\mbox{kHz}) +0.12 \,(f/1\mbox{kHz})^2,\qquad m=1.578;
$$
$$\mbox{EOS FPS}:\qquad
M \approx 1.4 +0.01 \,(f/1\mbox{kHz})^2 +0.0087 \,(f/1\mbox{kHz})^4,
$$
$$
j \approx 0.338 \,(f/1\mbox{kHz}) +0.19 \,(f/1\mbox{kHz})^2,\qquad m=1.56;
$$
$$\mbox{EOS L}:\qquad
M \approx 1.4 +0.019 \,(f/1\mbox{kHz})^2 +0.0345 \,(f/1\mbox{kHz})^4,
$$
$$
j \approx 0.56 \,(f/1\mbox{kHz}) +0.61 \,(f/1\mbox{kHz})^2,\qquad m=1.52;
$$
$$\mbox{EOS M}:\qquad
M \approx 1.4 +0.0195 \,(f/1\mbox{kHz})^2 + 0.071 \,(f/1\mbox{kHz})^4,
$$
\begin{equation}
j \approx 0.55 \,(f/1\mbox{kHz}) +1.04 \,(f/1\mbox{kHz})^2,\qquad m=1.494.
\end{equation}

The right-hand parts approximate $M(f),\quad j(f)$ in the entire range $0 \leq f<f_K$; they were constructed on the basis of tables from Cook et al. 
(1994). These approximations are extended to negative $f$ 
(counterrotation) by using the evenness condition for $M(f)$ and the oddness condition for $j(f):\quad j(-f)=-j(f)$. 

It follows from the formulas for $M(f)$ that the dependence of 
gravitational mass on angular velocity for the stiffer EOS L and M is 
stronger than that for the softer equations of state. While matching in 
the static limit, the gravitational masses of a NS with different 
equations of state differ at $f = 600 Hz$:
$$
M_A = 1.4034,\, M_{AU} =\,1.4043,\,M_{FPS}= 1.4047,\, M_L = 1.4113,\,M_M =
1.4162.
$$
The difference in the equations of state leads to a marked difference in 
the gravitational masses   (  for the same mass in the static limit  )   for 
the same rotation period. These differences exceed the accuracy of 
measuring the gravitational masses of millisecond pulsars in some binary 
systems (Thorsett and Chakrabarty 1998).

{\bf Determining the Equatorial Radius of a Neutron Star.} Another important 
formula relating the constant $\mu$ (stellar chemical potential) to the 
derivative of the gravitational mass with respect to the rest mass at 
constant angular momentum, follows from theorem 3 in SS 00. Taking the 
value of $\mu$ at the equator, we obtain
\begin{eqnarray}
&\mu = M_{,m}|_J= M,_{m}|_j/(1 + 2 j M,_{j}|_m)  = \sqrt{f - 2\omega f\tilde{\Omega} -(Rc^2/G M)^2\tilde{\Omega}^2},\\
& (R c^2/G M)^2 = t/f - f^2 \omega.
\end{eqnarray}
Here, $R$ is the NS geometric equatorial radius (the equator length 
divided by 2$\pi$), and  $\,\tilde{\Omega} = \Omega GM/c^3$ is its dimensionless angular velocity. Below, we use a stationary, axially symmetric metric in 
Papapetrou's form outside the star:
\begin{equation}
ds^2 = -F(dt - \omega d\phi)^2 + F^{-1}[ \rho^2 d\phi^2 + e^{2\gamma}/F (d\rho^2 + dz^2)]. 
\end{equation}
In the static limit, the stellar radius can be determined from formula 
(19) by using the Schwarzschild metric
\begin{equation}
\f{R_{st} c^2}{GM} = \f{2}{1 - M,_m^2},\qquad  M_{,m} \equiv \f{d M}{d m}.
\end{equation} 
Remarkably, formula (22) for a rotating NS, if $M(j, m)$ is substituted in it and differentiated at constant $j$:
\begin{equation}
R\approx \f{2G M/c^2}{1 - M,_m^2|_j},\qquad  M,_m|_j \equiv \f{\p M}{\p m}
\end{equation}
closely agrees with the numerical data for the stellar equatorial radius 
from Cook et al. (1994) and with our data obtained by using the 
numerical code of Stergioulas (1998) with an accuracy up to 1  \%.

{\bf The NS Equatorial Radius and the Gap between the Marginally Stable 
Orbit and the Equator.} Figure 5 shows plots of equatorial radius $R$ (in units of $GM/c^2$) against rotation frequency $f$ constructed using formula 
(23) for fixed gravitational masses $M = 1.2, 1.4, 1.6$, and 1.8$ M_{\odot}$ and for EOS FPS. In particular, the approximation of $R(f)$ at $M = 1.4 M_{\odot}$ has a fairly simple form:
\begin{equation}
R/1\mbox{km} = 11 + 1.78\, (f/1\,\mbox{kHz})^4.
\end{equation}
\begin{figure}[t]
\includegraphics[width=0.75\linewidth,bb=94 230 570 724,clip]{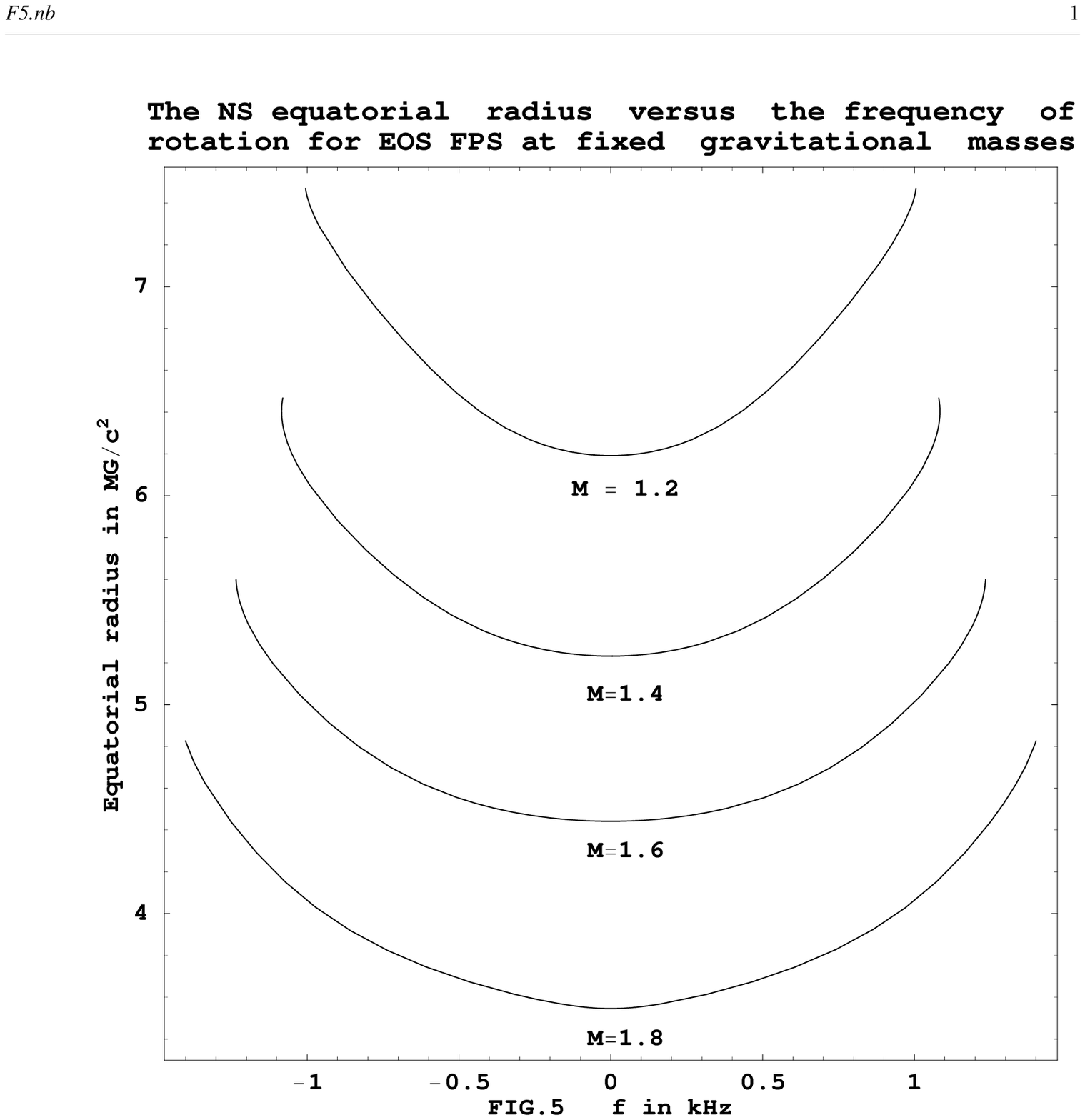}
\caption{\small Equatorial radius $R$ versus NS rotation frequency for fixed 
gravitational masses $M =$ const (EOS FPS). }
\end{figure}

Figure 6 shows plots of radius $R_*$ of the marginally stable orbit 
against $f$ constructed using formula (47) for the same values of $ M$. In 
particular, the approximation of $R_*(f)$ at $M = 1.4 M_{\odot}$ by a 
fourth-degree polynomial in the range $- 1\mbox{kHz} < f < 0.6$ kHz is (EOS FPS)
\begin{equation}
R_{*}/1\mbox{km}  = 12.44 -3.061\,( f/1\,\mbox{kHz})
+ 0.843\,(f/1\,\mbox{kHz})^2 + 0.6\,(f/1\,\mbox{kHz})^3 + 1.56\,(f/1\,\mbox{kHz})^4.
\end{equation}
\begin{figure}[h]
\includegraphics[width=0.75\linewidth,bb=89 283 515 724,clip]{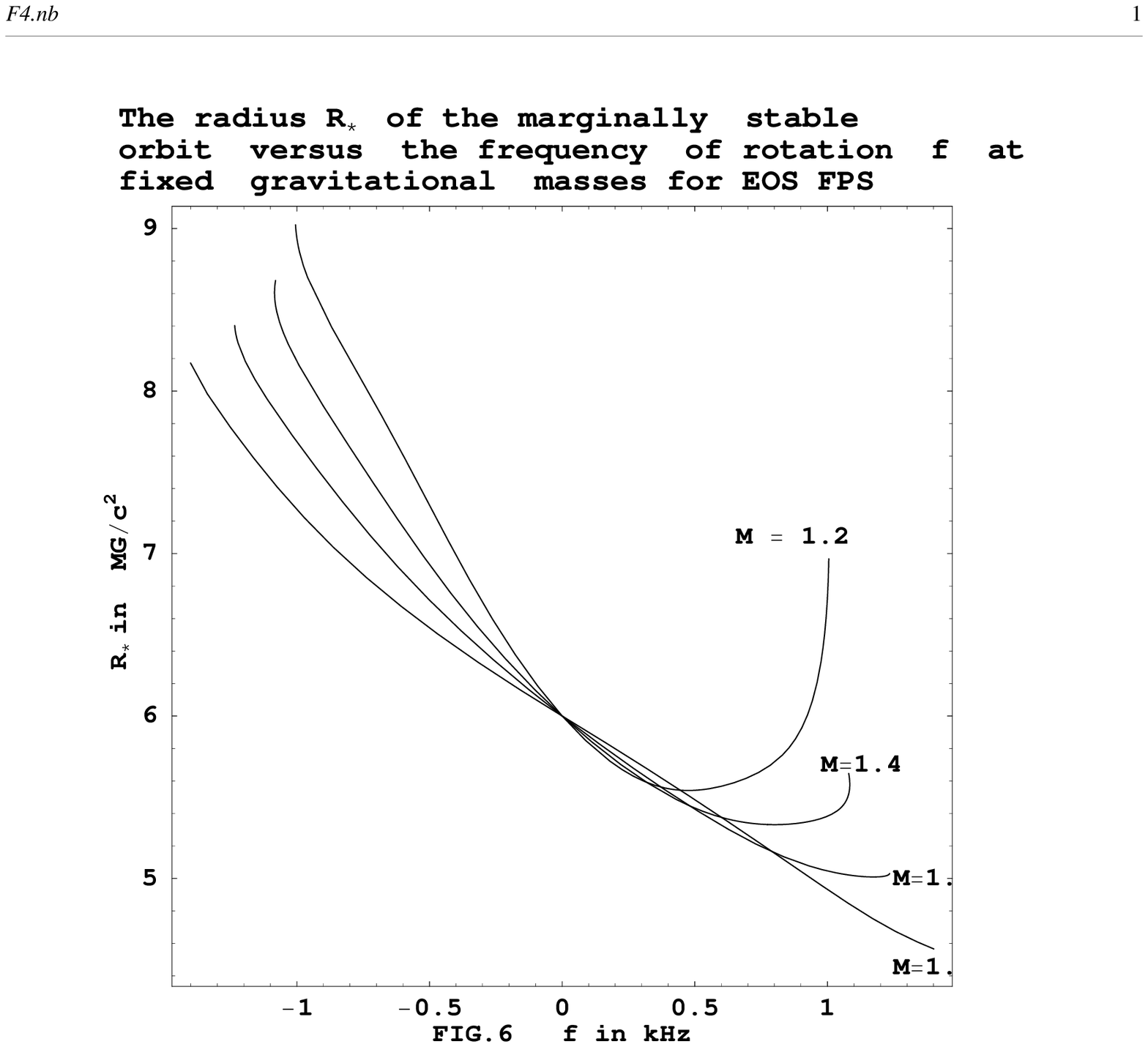}
\caption{\small Radius $R_*$ of the marginally stable orbit versus NS rotation 
frequency for fixed gravitational masses $M =$ const (EOS FPS). }
\end{figure}

To characterize the gap between the NS surface and the marginally stable 
orbit, let us consider the quantity $ R_*- R.$ In Fig. 7, $R_* - R$ is plotted 
against angular velocity for the same gravitational masses. The 
approximation of this dependence for $M = 1.4 M_{\odot}$ is
\begin{equation}
(R_* -R)/1\mbox{km}\approx 1.44 -3.061\,( f/1\,\mbox{kHz})
+ 0.843\,(f/1\,\mbox{kHz})^2 + 0.6\,(f/1\,\mbox{kHz})^3
 - 0.22\,(f/1\,\mbox{kHz})^4.
\end{equation}
These formulas can be used in the range $-1\mbox{kHz} < f < 0.6 kHz.$
\begin{figure}[h]
\includegraphics[width=0.85\linewidth,bb=11 250 570 716,clip]{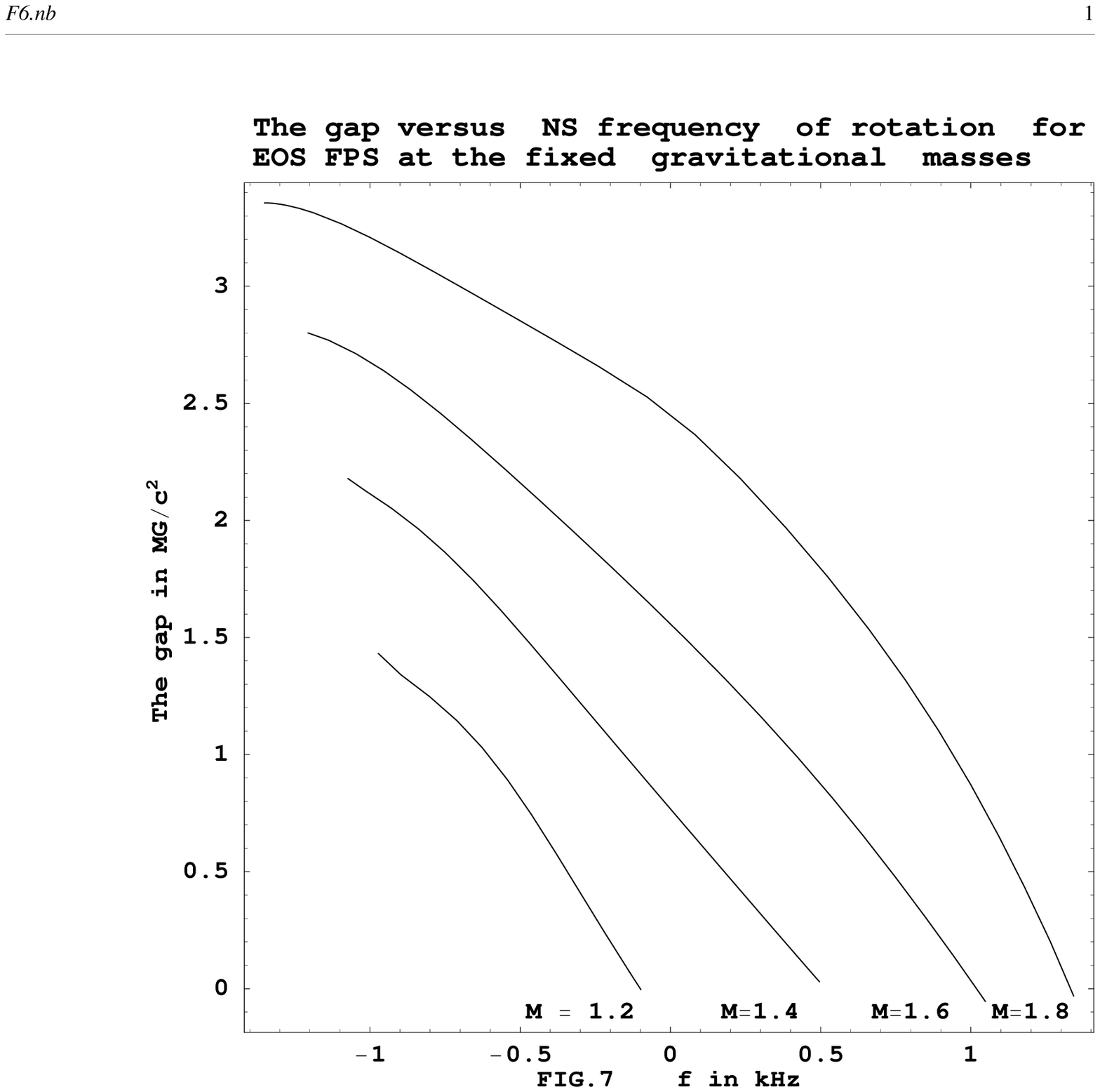}
\caption{\small   Depth $R_* - R$ of the gap between the marginally stable orbit and
the NS equator versus NS rotation frequency for fixed gravitational masses
$M =$ const (EOS FPS).  }
\end{figure}

In order to compare the gaps between the marginally stable orbit and the 
surface of a NS with different equations of state, we approximate the 
gap for $M = 1.4 M_{\odot}$ normal sequences for a NS with EOS A and EOS AU as 
follows:
\begin{equation}
\mbox{EOS A}:\qquad
(R_* -R)/1\mbox{km}\approx 3
-2.44\,f/1\,\mbox{kHz} - 0.2\,(f/1\,\mbox{kHz})^2.
\end{equation}
\begin{equation} 
\mbox{EOS AU}: \qquad
(R_* -R)/1\mbox{km}\approx 2.086
-2.81\,f/1\,\mbox{kHz} - 0.23\,(f/1\,\mbox{kHz})^2.
\end{equation}
\section{ The quadrupole moment of a rapidly rotating neutron star in
general relativity.}
To describe the oblateness effect and the emergence of a quadrupole 
moment in the mass distribution via rapid rotation, in SS 98 we proposed 
to use an exact quadrupole solution, which contains an arbitrary 
parameter $b$ compared to the Kerr metric. This parameter, to within a 
factor, matches the NS inherent quadrupole momentum: the total 
quadrupole moment in the asymptotics at large distances  $\Phi_2 = G^2 M^3/c^4 (b+j^2)$ 
includes the Kerr quadrupole moment. The parameter  $\Phi_2$ was determined by 
Ryan (1995, 1997)  with the use of  considerations developed by Komatsu et al. 
(1989) and Salgado et al. (1994).

{\bf The Metric of the Quadrupole Solution in the Equatorial Plane}. The 
metric components in the equatorial plane for this solution are (SS 98)
\begin{equation} 
F = \frac{A-B}{A+B},\quad \omega/M=-\frac{2jC}{A-B},\quad A\equiv
 (r_+ + r_-)^2r_+r_- - b,\quad B\equiv (r_+ + r_-)(r_+r_- - b+r).
\end{equation}
Here,
$$
C\equiv (r_+ + r_-)(r_-r_+ + r)+b,\quad
2r_{\pm}\equiv \sqrt{4r + (\sqrt{1-j^2}\pm\sqrt{1-j^2-4b})^2},\quad
r\equiv \rho^2/M^2.
$$
An exact quadrupole solution  of Einstein equations is contained as
a special case in the exact five-parameter solution found by Manko et 
al. (1994) by specifying its 
properties on the symmetry axis  using the method of Sibgatullin (1984). Ernst (1994) showed that this solution could also 
be obtained from Kramer - Neugebauer's (1980) solution for coaxially 
rotating black holes by a special choice of the constants in the 
solution.

The solution under consideration for $b < 0.25(1 - j^2)$ can be 
interpreted as a solution that describes the gravitational field of two 
coaxially rotating black holes with the same masses and angular momenta. 
For $b > 0.25(1 - j^2)$, it describes the gravitational field of coaxially rotating Kerr disks. In this case, the pressure clearly acts as elastic 
supports. For the curve of transition from black holes to Kerr disks to 
be found in the $j, m$ plane, the equation $ b(j, m) = 0.25(1 - j^2)$ must 
be solved.

{\bf Why the Kerr Metric Cannot Be Used to Describe the External Field of a 
Rapidly Rotating Neutron Star?}
 Let us consider the dependence of the NS 
quadrupole moment on its angular velocity for various equations of 
state, from the soft EOS A to the hard EOS L. We express the quadrupole 
moment $\Phi_2 = G^2 M^3 (b+j^2)/c^4 $ and the cyclic angular velocity in units of $10^{44} g cm^2$ and kHz, respectively. We fix the corresponding normal sequences 
(by definition, with a constant rest mass) by the condition that the NS 
gravitational mass is 1.4 $M_{\odot}$ in the static limit. Compare the total NS
quadrupole moment with the Kerr quadrupole moment $Q_K=G^2M^3 j^2/c^4$
(determined by the NS mass and angular momentum). Let us pass to an observable 
variable, the rotation frequency $f.$ Parabolic approximations of the 
dependence of dimensionless quadrupole coefficient $b$ on Kerr parameter $j$ are given in Laarakkers and Poisson (1998); more accurate 
approximations by fourth-degree polynomials in $j$ were derived in SS 98. 
Having reduced the data obtained with the numeric code of Stergioulas 
(1998), we have
$$
\mbox{EOS A}:\qquad\Phi_2 / 10^{44} g\,\mbox{ cm}^2 \approx 0.33 (f/1\,\mbox{kHz})^2 + 0.306 (f/1\,\mbox{kHz})^4,
$$
$$
Q_K / 10^{44} g\, \mbox{cm}^2 \approx 0.11 (f/1\,\mbox{kHz})^2 + 0.09 (f/1\,\mbox{kHz})^4;
$$
$$
\mbox{EOS AU}:
\qquad\Phi_2 / 10^{44} g\,\mbox{ cm}^2\approx 0.64 (f/1\,\mbox{kHz})^2 + 0.43 (f/1\,\mbox{kHz})^4,
$$
$$
Q_K / 10^{44} g\,\mbox{ cm}^2 \approx 0.16 (f/1\,\mbox{kHz})^2 + 0.12 (f/1\,\mbox{kHz})^4;
$$
$$
\mbox{EOS FPS}:
\qquad
\Phi_2 / 10^{44} g\,\mbox{ cm}^2 \approx 0.54 (f/1\,\mbox{kHz})^2 + 0.9 (f/1\,\mbox{kHz})^4,
$$
$$
Q_K / 10^{44} g\,\mbox{ cm}^2  \approx 0.13 (f/1\,\mbox{kHz})^2 + 0.22 (f/1\,\mbox{kHz})^4;
$$
$$
\mbox{EOS L}:
\qquad
\Phi_2 / 10^{44} g\,\mbox{ cm}^2 \approx 4.8 (f/1\,\mbox{kHz})^2 + 6.5 (f/1\,\mbox{kHz})^4,
$$
\begin{equation}
Q_K / 10^{44} g\, \mbox{cm}^2 \approx 0.54 (f/1\,\mbox{kHz})^2 + 1.16 (f/1\,\mbox{kHz})^4.
\end{equation}
Whereas the total quadrupole moment for the soft EOS A is approximately 
a factor of 3 larger than the Kerr component, for the hard EOS L it 
exceeds the Kerr component by almost a factor of 10!

{\bf Determining the Non-Kerr Quadrupole Moment of a Rapidly Rotating NS 
Using $M(j, m)$.} The NS equatorial radius $R$ is closely related to its 
quadrupole moment  $\Phi_2 = G^2 M^3 j^2(k+1)/c^4$. The system of equations (19) and (20) at 
given $j$ and $m$ in the metric (21) and (29) is algebraic for the NS 
coordinate radius $\rho$ and its quadrupole moment $b$, because, 
according to (9) and (23), the functions $\Omega(j, m)$ and $R(j, m)$ are 
expressed in terms of $M(j, m)$ and its partial derivatives. We emphasize 
that $M(j, m), \,R(j, m), \,b(j, m)$, and $I(j, m)$ (moment of inertia) are 
even functions of the Kerr parameter $j$, while the angular velocity 
$\Omega$ is an odd function of $j$.

In the approximation of slow rotation , $b \sim j^2$. Denote $b\equiv j^2 k$. The 
quadrupole moment can be approximated in finite ranges, $|j|\leq 0.67, 1.1< m <2.5$ , as a function of $j, \,m$ or as a function of $j, \,M.$

We derive an approximation formula for $k(j, m)$ by a method similar to 
that described above for constructing $M(j, m)$ and $m(j, M)$: the 
function $k$ is first approximated by a sixth-degree polynomial at the 
stability boundary in the segment $|j| < 0.67$ in terms of the rms 
deviation and then by eighth-degree polynomials in $j$ at two fixed 
values, say, $m = 1.56$ and 1.2358 ($M = 1.4$ and 1, respectively) in 
the segment $|j| < 0.67.$

{\bf The Quadrupole Moment as a Function of $j, $m or $j, M$ for EOS FPS.} 
The resulting formula for $k(j, m)$ is a combination of four rms 
approximations and describes $b(j, m)$ to within the third decimal place.

For the quadrupole moment of a NS with EOS FPS rotating arbitrarily fast 
(up to the Keplerian angular velocity at the stellar equator), the 
following formula holds:
\begin{eqnarray}
&\Phi_2 = G^2 M^3 j^2(k+1)/c^4,\qquad
 k \approx  k_*(j) +\nonumber\\
& (1.525 - 3.365\, j^2 - 9.138 \,j^4 + 23.37 \,j^6 - 28\, j^8)(m_*(j) -
m)^{0.5} +\nonumber\\
& (0.62 + 3.086\, (0.188 - 3.447\, j^2 + 4.83\, j^4+ 5.11\, j^6 -  19\, j^8)\nonumber\\
& (1.56 - m))\, (m_*(j) - m) +
 3.97\, (m_*(j) - m)^2\approx \nonumber\\
& k_*(j) + (1.693 - 4.196\, j^2 - 12.268\, j^4 + 30.32\, j^6 - 37.2\, j^8) 
\nonumber\\
 & (M_*(j) - M)^{0.5} + (1.713 + 2.5 \,(0.348 - 6.515 j^2 + 8.285 j^4
+\nonumber\\
& 9\, j^6 - 31\, j^8)\, (1.4 - M))\, (M_*(j) - M) + 5.453\, (M_*(j) - M)^2.
\end{eqnarray}
Here, $k_*(j)$ denotes the function at the stability boundary. It follows 
from our numerical data that $k_*(j)\approx 0.567 - 0.576\, |j| +
 0.338 \,j^2 - 0.47\, j^4 + 0.19\, j^6$;\, $M_*(j)$ and  $m_*(j)$  are given by
(16).

{\bf The Quadrupole Moment as a Function of $j, \,m$ for EOS A}. For EOS A, 
the formula for the inherent quadrupole coefficient can be written as
\begin{eqnarray}
&\Phi_2 = G^2 M^3 j^2\,(k+1)/c^4,\nonumber\\
k\approx & 0.556 - 0.862\, |j| + 0.795\, j^2-0.611\, j^4 +
(1.15 + 0.344\, |j|-3.68\, j^2-3.62\, j^4)\nonumber\\
&\sqrt{m_*(j) - m} + (1.196 +( 1.1075 -7.7013\, |j| +3.574\, j^2)\nonumber\\
&(1.5663 -
M_0))\,(m_*(j) - m) + 4.183\,(m_*(j) - m)^2.
\end{eqnarray}
Here, $M_*(j)$ and $m_*(j)$ are given by (14).

{\bf Results of Our Calculations.} Figure 8 shows lines of constant 
gravitational mass $M(f, m)$ as functions of rotation frequency $f$ for $m$ 
at 0.1 steps in the interval (1.2, 2.5) for EOS FPS (recall that the 
masses are measured in $M_{\odot}$). We used the parametric dependences $M = 
M(j, m)$ and $\Omega =\Omega(j, m)$ (see formula (9) for $\Omega(j, m)$ to 
construct these curves. The dashed curve in Fig. 8 separates the NS 
states when it is within the marginally stable orbit from the NS states 
when this orbit is inside it. Obviously, the equation of this curve is 
$R(j, m) = R_*(j, m)$. The dots indicate the curves of stability loss 
according to the static criterion. Their parametric equation is: $M = 
M_*(j)$ and $f = f_*(j).$
\begin{figure}[t]
\includegraphics[width=0.85\linewidth,bb=80 250 555 720,clip]{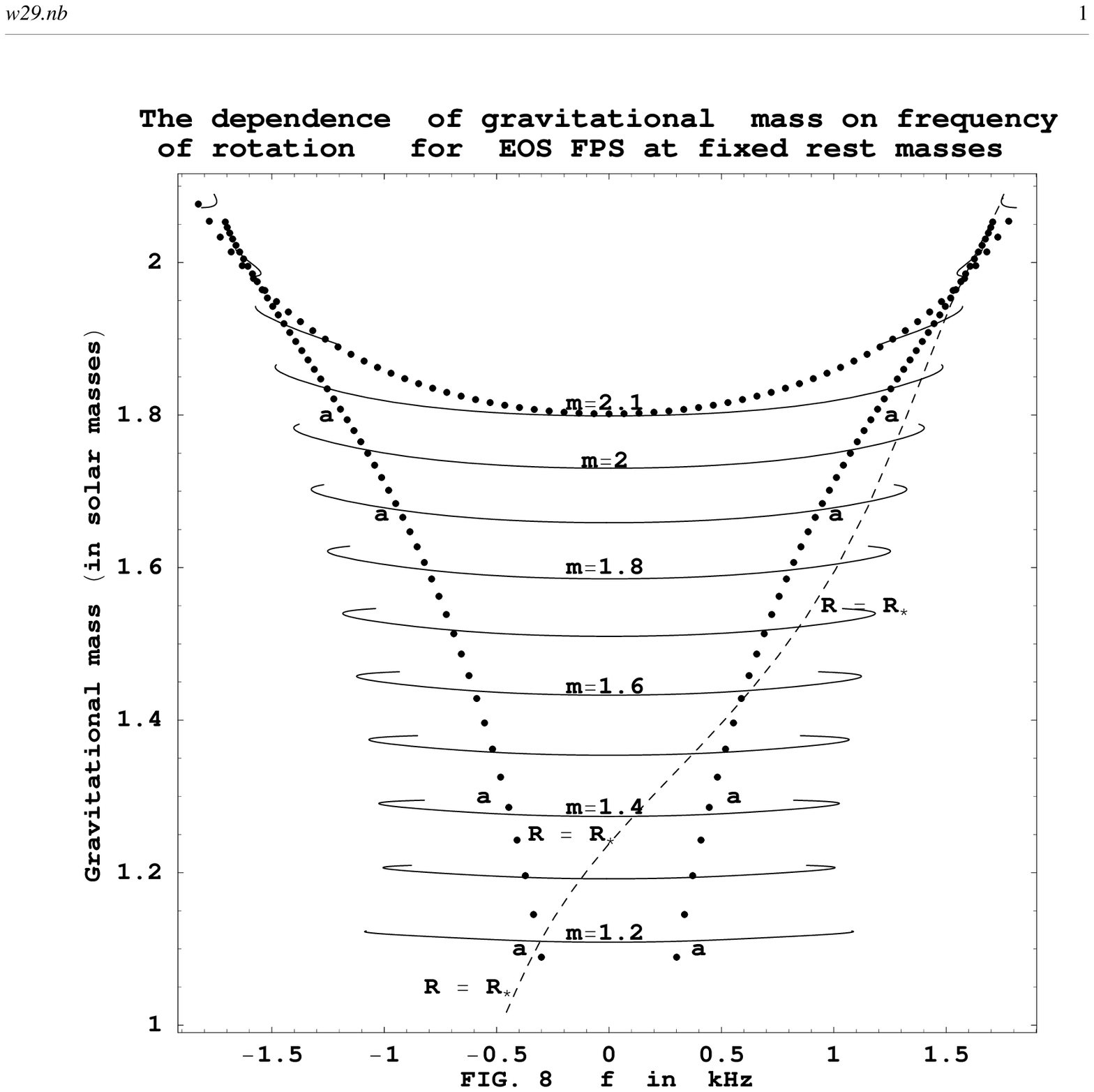}
\caption{\small  Gravitational mass $M$ versus NS rotation frequency for fixed rest
masses (EOS FPS). The dashed line corresponds to the parameters at which $R
= R_*.$ Curves $a,\,a,\,a \dots$ correspond to a transition of the external
field from the field of two Kerr black holes (inner region) to the field of
two Kerr disks (outer region). The maximum gravitational masses at which the
stability is lost according to the static criterion are also marked by dots
(upper curve).}
\end{figure}

The moment of inertia, the angular velocity, the equatorial radius, and 
the quadrupole moment can be inferred from the derived function $M(j,m)$, which determines the state of a two-parameter thermodynamic system.
\begin{figure}[t]
\includegraphics[width=0.85\linewidth,bb=94 230 560 724,clip]{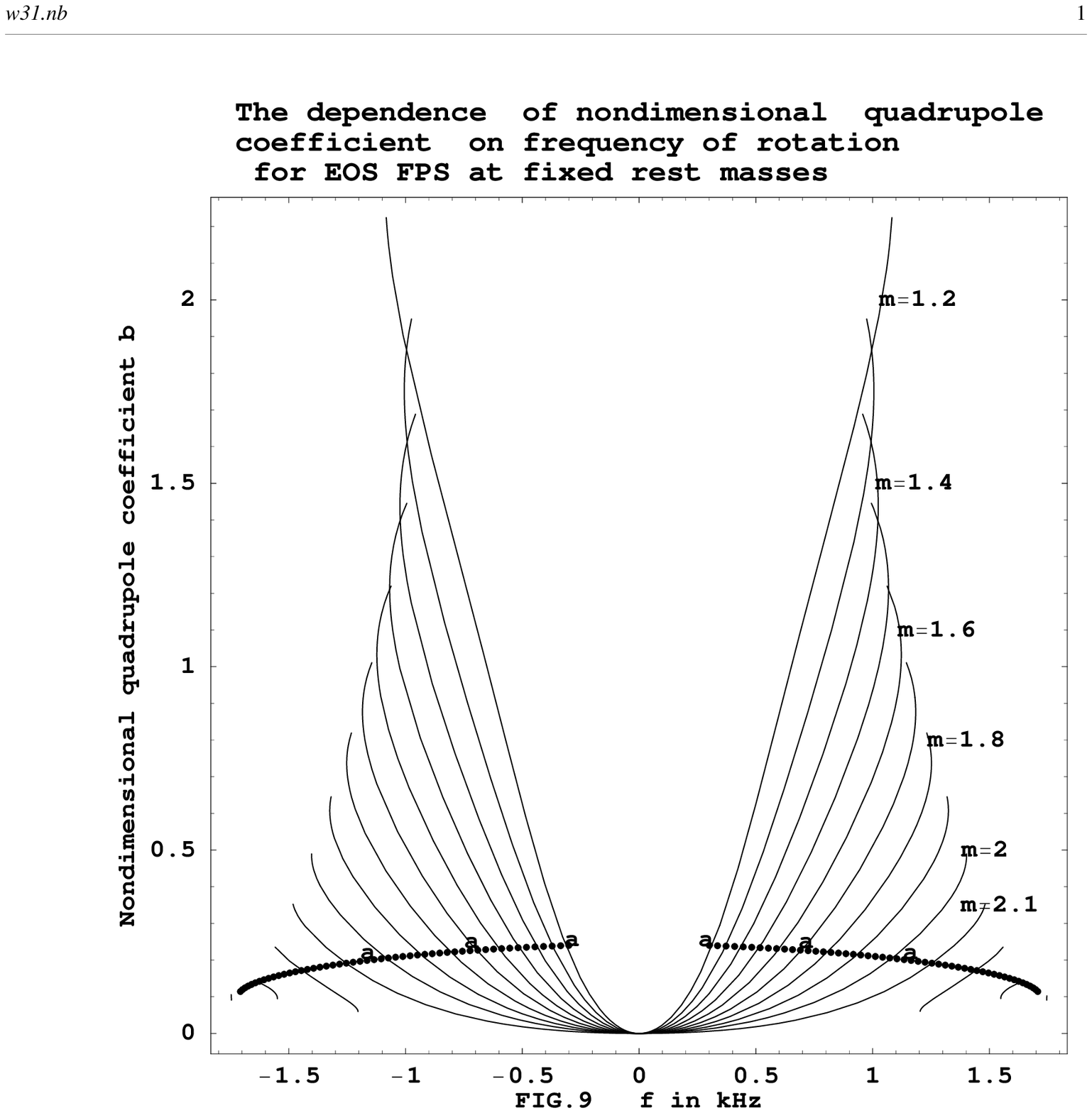}
\caption{\small Dimensionless quadrupole coefficient $b$ versus rotation frequency
$f$ at fixed rest masses $m$ (EOS FPS). Curves $a,\,a,\,a \dots$ correspond
to a transition of the external field from the field of two Kerr black holes
to the field of two Kerr disks. The curves were constructed at steps $ m =
0.1 M_{\odot}$.  }
\end{figure}
Figure 9 shows lines of constant dimensionless quadrupole coefficient $b$ 
as functions of rotation frequency $f$ for rest masses at 0.1 steps in 
the interval (1.2, 2.5). In this case, the dimensionless angular 
momentum $j$ acts as a parameter in the parametric specification of the 
curve on the plane. Points $a, \,a , \,a,$ ... in Figs. 8 and 9 correspond to 
the curve of transition from the external field of two coaxially 
rotating black holes to the external field of two coaxially rotating 
Kerr disks. The equation of this curve is $b(j, m) = 0.25(1 - j^2)$.
\section{The external gravitational fields of rapidly rotating neutron stars.}

{\bf The External Gravitational Fields of Sources with a Finite Set of 
Multipole Moments}. The external gravitational fields of rapidly 
rotating neutron stars at large angular velocities differ markedly from 
the Kerr field. To describe these fields by the solution of the Einstein 
equations with a finite set of multipole parameters, in SS 98 we 
proposed to use axisymmetric steady-state solutions specified on the 
symmetry axis by the following Ernst potential:
\begin{equation}
 e(z) = {\bf E}|_{\rho=0} = \frac{z^n-z^{n-1}+\sum_{j=1}^n a_j z^{n-j}}
 {z^n+z^{n-1} + \sum_{j=1}^n a_j z^{n-j}}.
\end{equation}
Below, the $\rho$ and $z$ coordinates are measured in units of length 
($GM/c^2$).

The corresponding solution is symmetric about the equatorial plane if we 
additionally require that the coefficients with even subscripts $a_{2k}$ be 
real (they are determined by the mass distribution and correspond to the 
Newtonian multipole moments), and that the coefficients with odd 
subscripts $a_{2k-1}$ be purely imaginary (they are determined by the 
angular-momentum distribution in the NS and have no analog in the 
Newtonian theory). For this definition of multipole moments, the Kerr 
solution is a purely dipole one, and its higher multipoles are zero. 
General expressions for the metric coefficients are given in SS 98 
[formulas (23) and (24)]. Denote the roots of the denominator in the 
expression for ${\bf E}$ on the symmetry axis by $ b_1,\, b_2, \dots, b_n$ and the 
roots of the equation $\tilde{e}(z) + e(z) = 0$ by   $\pm \xi_1, \pm\xi_2,
...,\pm \xi_n$ (Sibgatullin 1984). Here, the tilde has the meaning of a complex
conjugate. If we use the identity
$$
\prod_{k=1}^{n} P(a_k)\det{\left(\f1{a_i-b_j}\right)} = -V(a_1,...,a_n)V(b_1,...,b_n)
$$
here,$ V(a_1, \dots, a_n) $ is an $n\times n$ alternant on elements  $a_1,...,a_n$, $\quad P(z) $  is an $n-th$ degree polynomial whose roots are $b_1, \dots,
b_n$, then we can represent the solution for the Ernst function 
differently. To describe it, we denote
\begin{equation}
\gamma_k \equiv a_1 + \f{a_3}{\xi_k^2}+ ...; \quad \sigma_k^{\pm}\equiv \pm\sqrt{1-\gamma_k^2} + i \gamma_k;\quad r_k^{\pm}\equiv\sqrt{\rho^2 + (z \mp \xi_k)^2};
\end{equation}
\begin{equation}
a_{kl}\equiv  r_l^{+}\sigma_l^{+}\xi_l^{k-1};\quad b_{kl}\equiv  r_l^{-}\sigma_l^{-}(-\xi_l)^{k-1};\quad c_{kl}\equiv \xi_l^{k-1};\quad d_{kl}\equiv (-\xi_l)^{k-1}.
\end{equation}
The solution for the Ernst function symmetric about the equatorial plane 
and with the specified behavior on the symmetry axis (28) then takes a 
fairly elegant form:
\begin{equation}
{\bf E} = \f{U - W}{U + W};\quad U = \det{\left(\begin{array}{cc}A,& B\\C,& D\end{array}\right)},
\quad W = \det{\left(\begin{array}{cc}A^{\prime},& B^{\prime}\\C^{\prime},& D^{\prime}\end{array}\right)}.
\end{equation}
The square $n \times n$ matrices $A, B, C, D$ consist of $a_{kl},\, b_{kl},
\,c_{kl}, \,d_{kl},\quad k,l = 1,...,n $, respectively [see Eq. (35)]. The
rectangular $(n - 1)\times n$ matrices $A^{\prime}, B^{\prime}$ consist of $a_{kl},\, b_{kl}, \quad k = 1,...,n-1;\,l = 1,...,n $ , and the rectangular $(n + 1) \times n$ matrices $C^{\prime}, 
D^{\prime}$ consist of $c_{kl}, \,d_{kl},\quad k=1,...,n+1;\,l =1,...n $.

Formally, the solution (36) appears as the result of applying Backlund's 
transformation to the solution ${\bf E} = 1$  $2 n$ times. For the Ernst 
equation, it was found by Neugebauer (1980); see also formula (7) from Kramer 
and Neugebauer (1980) with ``Van der Mond representation'' of Ernst function. However, an attempt to directly determine the 
parameters of this solution from the data on the axis leads to 
cumbersome calculations even for $n =2$. Ernst
1994 established the relation between Sibgatullin's
method of construction of electrovac solutions with the given rational
behavior at the symmetrie axes and Neugebauer's 2n parametric family of
solutions for $n = 2.$ Manko and Ruiz 
(1998) were able to represent the solution of the Ernst equations 
corresponding to the Ernst rational function on the symmetry axis with 
the asymptotics ${\bf E} \rightarrow 1$  when $z\rightarrow \infty$ in a form that contained only the 
roots $\xi_k$ of the equation $\tilde{e}(z) + e(z) = 0$ and that did not 
contain the roots $b_k$  for arbitrary $n>2.$

We hypothesized in SS 98 that the coefficients $a_{2k-1}$ for rigidly 
rotating stars, which are related to differential rotation, were zero 
for $k > 1$, with $a_1 = j \ne 0.$ This coefficient is the ratio of the NS 
angular momentum to its mass squared. Then, a substantial simplification 
of the solution (36) is that all constants $\,\sigma_k^{+}$ and  $\,\sigma_k^{-}\,$ (defined in (34)) turn out to be equal:
$$\sigma_k^{+}=\sqrt{1-j^2} + ij,
 \quad\sigma_k^{-} = -\sqrt{1-j^2} + ij,\quad k = 1,...,n.$$

{\bf An Exact Solution of the Einstein Equations for a Rotating Deformed 
Source and Numerical Data}. The available numerical data on the external 
gravitational fields of rigidly rotating neutron stars [redshifts forward $Z_f$
and backward $Z_b$  at the equator edges, which allow the 
metric coefficients $F$ and $\omega$ on the stellar equator to be calculated; 
see Cook et al. (1994) for numerical values of the radius of the
marginally stable orbit] suggest that these fields can be described by 
some exact solution of the Einstein equations. The corresponding Ernst 
function on the symmetry axis is
\begin{equation} 
e(z) = \f{z^2 - (1 - ij)z +b}{z^2 + (1 + ij)z + b}.
\end{equation}
which is obtained from the general case (33) at $n = 2.$  Recall that $J = GM^2j/c$ is the angular 
momentum, and  $G^2 b M^3/c^4$ is the quadrupole moment of the NS. Manko et 
al. (1994, 2000) proposed to model the external fields of neutron stars 
with strong magnetic fields by special exact solutions of the system of Einstein-Maxwell equations.

In contrast to the multipole decompositions at large radii (Shibata and 
Sasaki 1998; Laarakkers and Poisson 1998), which converge slowly at the 
stellar surface, the quadrupole solution closely approximates numerical 
data up to the stellar surface. The parameter $b = b(j)$ of this 
solution can be independently determined by several methods: for 
example, by comparing either the radius of the marginally stable orbit 
or the metric coefficient $F$ at the stellar surface, which is
$((1+Z_f)\,(1+Z_b))^{-1}$ (recall that $ 1/\sqrt{F}-1$ has the meaning of gravitational 
redshift), with numerical data. The metric coefficients $F$ and $\omega$ in the
solution corresponding to (37) on the symmetry axis take the form (29) 
in the equatorial plane.

Remarkably, $b = b(j)$ determined from independent comparisons with 
numerical data proved to be the same for a given equation of state and 
at fixed rest mass.

The function $R_*(j,m)c^2/G M(j,m)$ at $m=\mbox{const}$ constructed from the Kerr solution differs 
markedly from the realistic curves for $|j| > 0.15.$ Nevertheless, the 
realistic $R_*\,c^2/G M$ curves and the Kerr curve have a tangency of the first order at $j = 0.$ This circumstance serves as a good illustration of 
the remarkable observation by Hartle and Thorne (1969) that the external 
gravitational field of a slowly rotating star is described by the Kerr 
metric linearized in rotation parameter.

Here, by contrast to SS 98, we approximated the function $b(j, m)$ in 
finite ranges of $j$ and $m$ [see formulas (26) and (27)].
\section{ Global properties of the exact quadrupole solution.} 
{\bf Parameters of Equatorial Circular Orbits in an Arbitrary Axisymmetric 
Stationary Field in a Vacuum}. The specific energy and angular momentum 
of particles rotating in the equatorial plane in a Keplerian circular 
orbit in an arbitrary axisymmetric stationary field in a vacuum can be 
calculated using the formulas (SS 98)
\begin{equation}
E=\frac{\sqrt{F}}{\sqrt{1-F^2 p^2/r}}, \quad l=MG/c (p+\omega)E,
\end{equation}
$$p\equiv r(-\lambda +\sqrt{\lambda^2+\mu-\mu^2r})/n,\quad \lambda\equiv F\dot{\omega},\,\mu\equiv
\dot{F}/F,\,n\equiv F-r\dot{F}.
$$
Here, the dot denotes a derivative with respect to $r\equiv \rho^2$.

For the angular velocity of a particle in a Keplerian circular orbit, we 
can derive the formula
\begin{equation}
\Omega_K =\f{c^3}{G M}\f{p}{(r/F^2 + \omega p)}.
\end{equation}
The radius of the marginally stable orbit can be determined by using the 
condition of energy extremum in circular orbits. In explicit form, it 
appears as
$$
r(\lambda \mu n+\dot{\lambda} n-\dot{n}\lambda )\sqrt{\lambda^2+\mu-\mu^2
r}-(\lambda^2+\mu-\mu^2 r)(n-r\dot{n})-
$$
\begin{equation}
rn(\lambda\dot{\lambda}+0.5(\dot{\mu}-\mu^2)-r\mu\dot{\mu})=0.
\end{equation}
For the particle energy and angular momentum in the marginally stable 
orbit to be calculated as functions of rotation parameter $j$, the root 
of the algebraic equation (40) must be substituted in (38).

{\bf Disk Luminosity and Parameters of Equatorial Circular Orbits in the 
Kerr Field}. The external field of a slowly rotating NS can be described 
by the Kerr solution linearized in angular momentum (Hartle and Thorne 
1969). The physical processes in the field of a slowly rotating NS were 
considered by Kluzniak and Wagoner (1985), Sunyaev and Shakura (1986), 
Ebisawa et al.(1991), Biehle and Blanford (1993), and Miller and Lamb 
(1996).

For the extreme Kerr solution ($j = 1$), the particle energy and 
angular momentum in the marginally stable orbit were calculated by 
Ruffini and Wheeler (1970) in their pioneering study.

At $0 < j < 1$, the particle energy and angular momentum in circular 
orbits in the Kerr field (Bardeen et al. 1972) are related to the 
orbital radius and the rotation parameter $j$ by
\begin{equation}
E = \frac{1 - 2x + j x\sqrt{x}}{\sqrt{1 - 3x + 2j x\sqrt x}},\quad
l =G M/c\frac{-2j x +(1 + j^2 x^2)/\sqrt{x}}{\sqrt{1 - 3x + 2j x\sqrt x}},
\quad x\equiv  G M/(r c^2).
\end{equation}
Here, $r$ is the Boyer - Lindquist radial coordinate in the Kerr metric.

The expressions for $j, E, l$  corresponding to the marginally 
stable orbit, where the $x$  is treated as a parameter, are
\begin{equation}
\label{f2}
j_* = \frac{4\sqrt{x} - \sqrt{3 - 2x}}{3x},\quad E_* = \sqrt{1 -
\frac23 x},\quad c l_*/GM - j_* E_* = \frac{1}{x\sqrt{3}}.
\end{equation}
The parameter $x$ varies in the intervals (1/9, 1/6) and (1/6, 1) for 
disk and black-hole counterrotation and corotation, respectively. The 
corresponding values of $j$ for the former interval are negative.

The disk energy release in the field of a black hole is obviously 
$ \dot{M} c^2(1 - E_*).$  In order to express the energy release as a function of the
black-hole angular velocity $\Omega$, we use the formula [formula (12) in Christodolou and Ruffini (1973); see also Misner et al. (1973), 
Sibgatullin (1984), Novikov and Frolov (1986)]
\begin{equation}
2\tilde{\Omega} =j/(1+\sqrt{1-j^2}),\quad \tilde{\Omega}\equiv\Omega M G/c^3.
\end{equation}
Hence, it is easy to obtain
$$
j =\f{4\tilde{\Omega}}{(4\tilde{\Omega}^2 +1)}.
$$
Therefore, the energy release $L_d$ in the disk around a Kerr black hole 
is a function of  $\tilde{\Omega}=\Omega M G/c^3$ specified parametrically with $x =G M/r c^2$  
 ($r$ is the Boyer-Lindquist coordinate radius of the marginally 
stable orbit). $L_d$ has the following well-known values: $L_d = 
0.0377 M c_2$ at $\tilde{\Omega} = - 0.5$  ($j=-1$); $L_d = 0.0572Mc^2 $ at
$\tilde{\Omega} = 0$  ($j=0$)); and $L_d = 0.4226Mc^2$ at $\tilde{\Omega}=0.5$  ($j=1$). Note that $j=1$ corresponds to a rotation frequency $f 
= c^3/(2\pi G M)\approx 23.1$ kHz for a black hole with $M=1.4 M_{\odot}.$ The Taylor 
expansion of the energy release in the disk around a Kerr black hole in 
the interval $-0.4\leq \tilde{\Omega}\leq 0.4\quad   ( -0.975 < j < 0.975)$  is
\begin{equation}
L_d /\dot{M} c^2\approx 0.0572 + 0.128\,\tilde{\Omega}+0.349\,\tilde{\Omega}^2
+0.532\,\tilde{\Omega}^3+0.52\,\tilde{\Omega}^4 + 0.417\,\tilde{\Omega}^5+0.65\,\tilde{\Omega}^6.\end{equation}
The function $L_d(\tilde{\Omega})/\dot{M} c^2 $  is not analytic at $\tilde{\Omega}=0.5.$ Its expansion in 
terms of fractional powers of the difference  $0.5  -\tilde{\Omega} $ is
\begin{equation}
L_d /\dot{M} c^2\approx 0.4226 - 1.155 (0.5  -\tilde{\Omega})^{2/3} + 1.443 (0.5
-\tilde{\Omega})^{4/3} - 0.77(0.5  -\tilde{\Omega})^{5/3} - 0.5(0.5  -\tilde{\Omega})^2 .\tag{$44a$}
\end{equation}
The calculation using formula (44) at  $\tilde{\Omega}=0.4$ ($j\approx 0.975$) 
yields 0.2185; this value differs from that calculated by using formula 
(44a) and from the exact value by less than 0.0005. One should therefore use
(44) for $|\tilde{\Omega}|<0.4$   and the expansion (44a) for $0.4
<\tilde{\Omega}\leq 0.5.$  In the 
range $-0.5
<\tilde{\Omega}\leq 0.4
$, the first two terms in the Taylor expansion 
near $\tilde{\Omega} \approx -0.5 (j \approx -1)$ can be used for the Kerr-disk luminosity:
$$
L_d/\dot{M}c^2 \approx 0.0377 + 0.024(\tilde{\Omega} + 0.5)^2.
$$
Note that, if the NS radius is smaller than the radius of the marginally 
stable orbit, then the energy release $L_d$ in the disk around the NS is 
approximately equal to $L_d$ in the disk around a black hole of the same 
mass and the same angular momentum. Indeed, we see from formula (45) for 
$E_*$ that the corrections to the Kerr expression associated with powers of 
the quadrupole coefficient $b$ are small.

{\bf Comparison of the Functions $j(\tilde{\Omega})$ for Black Holes and Neutron Stars.} As 
the NS mass increases to its maximum value, when the star collapses in 
the static limit, the relationship between the dimensionless angular 
velocity of a NS with different equations of state and the Kerr 
parameter approaches the above relationship between these parameters for 
rotating black holes. Moreover, the NS external field differs only 
slightly from the gravitational field of a rotating black hole.

Indeed, using tables from Cook et al. (1994), we can construct 
approximations for the $M = 1.4 M_{\odot}$ normal sequences in the static limit 
in the range $0 < j < 0.6$ [the function $j(\tilde{\Omega})$ is extended to negative 
$\tilde{\Omega}$ by using the oddness condition $j(\tilde{\Omega}) =-j(\tilde{-\Omega})$]:
$$
\mbox{EOS A}\qquad j\approx 6.573
\tilde{\Omega}+0.655(10\tilde{\Omega})^2,\quad m=1.566;
$$
$$
\mbox{EOS AU}\qquad j\approx 8.2
\tilde{\Omega}+0.65 (10\tilde{\Omega})^2,\quad m=1.577;
$$
$$
\mbox{EOS FPS}\qquad j\approx 8.035
\tilde{\Omega}+0.91 (10\tilde{\Omega})^2,\quad m=1.56;
$$
$$
\mbox{EOS L}\qquad j\approx 13.3
\tilde{\Omega}+3.03 (10\tilde{\Omega})^2,\quad m=1.52.
$$
Let us now consider the functions $j(\tilde{\Omega})$ of maximum masses stable only in 
the presence of rotation:
$$
\mbox{EOS A}\quad j\approx 4.066
\tilde{\Omega}+0.388(10\tilde{\Omega})^2,\quad m=1.92;\quad 0\leq 10\tilde{\Omega}
\leq 0.84
$$
$$
\mbox{EOS AU}\quad j\approx 4.095
\tilde{\Omega}+0.14 (10\tilde{\Omega})^2,\quad m=2.638;\quad 0\leq 10\tilde{\Omega}
\leq 1.1
$$
$$
\mbox{EOS FPS}\quad j\approx 4
\tilde{\Omega}+0.445 (10\tilde{\Omega})^2,\quad m=2.1;\quad 0\leq 10\tilde{\Omega}
\leq 0.8;
$$
$$
\mbox{EOS L}\qquad j\approx 4.33
\tilde{\Omega}+0.34 (10\tilde{\Omega})^2,\quad m=3.23 \quad 0\leq 10\tilde{\Omega}
\leq 0.86
$$

It follows from the above formulas that, for the maximum possible 
masses, the functions $j(\tilde{\Omega})$ of arbitrary NS equations of state are 
closely approximated by the dependence for Kerr black holes (43) at 
small $\tilde{\Omega}$:$j\approx 4\tilde{\Omega}$. At the same time, $j(\tilde{\Omega})$ for neutron stars with  $M = 1.4 M_{\odot}$ have  $j(\Omega)$  considerably exceed the Kerr values.

{\bf Parameters of the Marginally Stable Orbit in the Equatorial Plane in 
the Field of a Rotating NS.} The external fields of rapidly rotating 
neutron stars differ markedly from the Kerr field. This difference can 
be described by introducing only one multipole moment, namely, the 
quadrupole one (Laarakkers and Poisson 1998; SS 98) at stellar masses 
larger than $M_{\odot}$. In general, the external gravitational fields at  $ M <M_{\odot}$ in the case of rapid rotation have all multipole components, much like 
the external field of a Maclaurin spheroid at a rotation velocity 
comparable in magnitude to the Keplerian velocity on the stellar 
equator.

In order to calculate the parameters of the marginally stable orbit, let 
us consider four quadrupole coefficients: $b: b = 0,\, 0.25,\, 0.5,\, 0.75$. 
For each of these values, we break up the $j$   range -0.7 to +0.7 into 30 
equal parts and seek a minimum of the particle energy in circular orbits 
 for each $j$ [rather than seek the root of the complex equation (40), as 
we did in SS 98]. Having determined the corresponding radial 
coordinates, we can calculate all the remaining parameters of the 
marginally stable orbit and construct an approximating polynomial in $j$ 
with the smallest rms deviation in the interval (-0.7, 0.7) at fixed 
$b.$ We then construct an interpolation polynomial in $b$ using 
Lagrange - Silvester's formula. Let us write out the functions (38) and 
(39) derived in this way, which, however, have a meaning only when the 
stellar radius is smaller than the radius of the marginally stable 
orbit. In the formulas given below, $ 1 - E_*$ has the meaning of binding 
energy of a particle of unit mass in the marginally stable orbit, and 
$\Omega_*$ is the angular velocity in this orbit:
\begin{eqnarray}
E_*\equiv & \min E = 
0.943 - 0.031\, j - 
    0.022\, j^2 - 0.014\, j^3 - 
    -0.01\, j^4 - 0.02\, j^5 -0.02\, j^6+ \nonumber\\
  &  b\,(0.008 + 
          0.017\, j + 0.024\, j^2 + 
          0.016\, j^3 + 0.013\, j^4 + 0.085\,j^5+ 0.1\, j^6) +\nonumber\\ 
   & b^2\,(-0.002 - 0.01\, j - 
          0.021\, j^2 - 0.005\, j^3 - 
          0.002\, j^4 - 0.14\, j^5-0.18\,j^6) +\nonumber\\  
   & b^3\,(0.001 + 
          0.004\, j + 0.009\, j^2 - 
          0.002\, j^3 - 0.004\, j^4+ 0.08\, j^5 + 0.11\, j^6);\\
c l_*/ (G M) = &3.464 - 0.943\, j - 
     0.258\, j^2 -0.125\, j^3 - 
      0.074\, j^4  - 0.11\, j^5- 
  0.09\,j^6 +\nonumber\\ 
     & b\, (0.189 + 
            0.226\, j + 0.244\, j^2  + 
     0.178\, j^3 + 0.15\, j^4+ 
    0.49\, j^5+ 0.51\, j^6)+\nonumber\\ 
     & b^2\, (-0.039- 0.108\, j - 0.191\, j^2- 
     0.123\, j^3  - 0.123\, j^4- 
     0.85\, j^5- 0.96\,j^6) + \nonumber\\ 
     & b^3
\, (0.009 + 
           0.036 \, j + 0.078\, j^2 + 
            0.034\, j^3 +0.036\, j^4 +0.5\, j^5+ 0.59\, j^6); \\
R_* c^2/GM  = & 6 - 3.267\, j - 
    0.278\, j^2 - 0.112\, j^3 - 
   0.059\, j^4  - 0.07\, j^5 - 
  0.05\, j^6+\nonumber\\
 & b\,(1.085 + 
          0.96\, j +0.904\, j^2 + 
          0.723\, j^3 +0.177\, j^4 +0.63 \,j^5 + 1.53\, j^6)+\nonumber\\
  &  b^2\,(-0.25 - 0.607\, j - 1.134\, j^2 - 
     1.107\, j^3 + 0.643 \,j^4 - 1.18\, j^5 - 
     4.89\, j^6) + \nonumber\\
  &  b^3\,(0.059 + 
         0.223\, j + 0.633\, j^2 + 
         0.604\, j^3- 0.923\, j^4 + 
     0.71\, j^5+ 3.94\, j^6);\\
\Omega_K^* G M/c^3 =& 0.068 + 0.051\, j + 
    0.037\, j ^2 + 0.025\, j^3 + 
  0.017\, j^4 +0.03\, j^5+0.03\, j^6+  \nonumber\\  
  &  b\,(-0.016 - 0.035\, j- 0.052\, j^2 - 
     0.039\, j^3- 0.033\, j^4 - 
     0.16\, j^5 - 0.18\, j^6) +  \nonumber\\ 
  &  b^2\,(0.006 + 
         0.024\,  j + 0.053\, j^2 + 
          0.028\, j^3 +0.014\, j^4+ 0.29\,j^5 +0.37\, j^6) + \nonumber\\ 
&  b^3\,(-0.002 - 0.01\, j - 
          0.025 \,j^2 - 0.007\, j^3 + 0.008\, j^4 - 0.18\, j^5 - 
     0.24\, j^6). 
\end{eqnarray}
These formulas are universal and valid for any NS equation of state. If, 
alternatively, the dimensionless quadrupole coefficient $b = k j^2$ is 
expressed in terms of $j$ and $m$ for a specific equation of state 
(respectively, $j$ and $M$), for example, by using formula (31), then we 
obtain the above formulas as functions of $j$ and $m$ (respectively, $j$ and $M$). Formula (32) must be used for EOS A.

We emphasize that, in contrast to the Taylor expansions of Shibata and 
Sasaki (1998) for small $j$ and $b\sim j^2$ , formulas (45 - 48), which were 
derived from the exact solution by the combination of least squares and 
interpolation in $b$ (Silvester-Lagrange's formula), describe the 
behavior of the solution in finite ranges: $b\leq 0.75, |j|\leq 0.7$. Note 
that the corrections associated with the coefficient $b$  in formula (45) for $E_*$ are small.

{\bf Radial and Azimuthal Velocities on the NS Surface for the Particles 
Falling to the Stellar Equator from the Marginally Stable Orbit.} The 
radial and azimuthal velocities of the particles that fell from the 
marginally stable orbit on the stellar surface in a local frame which is 
stationary relative to the orbits of Killing's time-like vector of the 
metric (21) are
$$
\f{V_{\phi}}{c} = \f{f}{\sqrt{r}}(p_* +\omega_* -\omega),\quad
\f{V_r}{c}=\sqrt{1-\f{f^2}{r}(p_* +\omega_* -\omega)^2-(1-\f{f_*^2
p_*^2}{r_*})\f{f}{f_*}}.
$$
Here, the asterisk denotes the corresponding parameter in the marginally 
stable orbit. The approximations for $V_{\phi}$ and $\,V_r$ for a NS with fixed 
mass $M=1.4 M_{\odot}$ and EOS A valid in the range $ -1.2 \mbox{kHz}< f<
0.88\mbox{kHz}$ are
$$
\f{V_{\phi}}{c}\approx 0.6 -0.006 (f/1\,\mbox{kHz})-
0.04\,(f/1\,\mbox{kHz})^2,\quad \f{V_r}{c}\approx 0.062 -0.057 (f/1\,\mbox{kHz})-
0.015\,(f/1\,\mbox{kHz})^2,
$$
and those for EOS FPS valid in the range $ -1. \mbox{kHz}< f<
0.6\mbox{kHz}$ are:
$$
\f{V_{\phi}}{c}\approx 0.553 -0.018 (f/1\,\mbox{ËçÃ})-
0.052\,(f/1\,\mbox{ËçÃ})^2,\quad \f{V_r}{c}\approx 0.03 -0.044 (f/1\,\mbox{kHz}) - 0.03\,(f/1\,\mbox{kHz})^2.
$$
In SS 98, we gave formulas for $V_{\phi}$ and $\,V_r$ in a frame
entrained by 
NS rotation. The corresponding approximations for $M=1.4 M_{\odot}$ for the 
azimuthal velocity are
$$
\mbox{EOS A}:\qquad \f{V_{\phi}}{c}\approx 0.6 - 0.035 (f/1\,\mbox{kHz})-0.042  (f/1\,\mbox{kHz})^2,
$$
$$
\mbox{EOS FPS}:\qquad \f{V_{\phi}}{c}\approx 0.553 -0.046 (f/1\,\mbox{kHz})-
0.055\,(f/1\,\mbox{kHz})^2.
$$
The radial components of the 4-velocity do not change when passing from 
one frame to the other.

The kinetic energy of radial particle motion produces energy release in 
a shock wave near the equator, while the energy of azimuthal motion 
produces energy release in the spread layer that concentrates around two 
bright latitudinal rings (Inogamov and Sunyaev 1999). Clearly, this 
difference can, in principle, allow the case with a radius smaller than 
the radius of the marginally stable orbit  $R\,<\,R_*$  (three bright rings) 
to be experimentally distinguished from the case with $R\,>\,R_*$(two 
bright rings).

Since the particles in the gap are assumed to have no time to gather a 
high radial velocity for any reasonable $f \quad V_r << V_{\phi}.$ Expanding 
the fourth tetrad 4-velocity component in a series yields
$$
\f{1}{\sqrt{1 - V_r^2/c^2 - V_{\phi}^2/c^2}}\approx \f{1}{\sqrt{1-V_{\phi}^2/c^2}}+\f{V_r^2/c^2}{\sqrt{(1 -V_{\phi}^2/c^2)^3}}.
$$
The flux of radial kinetic energy is therefore
$$
L_{rad}\approx \dot{M}\f{V_r^2/2}{\sqrt{(1 -V_{\phi}^2/c^2)^3}}.
$$
The approximation of $L_{rad}$ for a NS with the soft EOS A, where this flux 
reaches a maximum, is
$$
L_{rad}/\dot{M}c^2 \approx 0.0035 -0.01(f/1\,\mbox{kHz}) + 0.0025(f/1\,\mbox{kHz})^2+0.005(f/1\,\mbox{kHz})^3.
$$
For the stiffer EOS FPS, the flux of radial energy is considerably lower,
$$
L_{rad}/\dot{M}c^2 \approx 0.00026 -0.003(f/1\,\mbox{kHz}) +
0.0036(f/1\,\mbox{kHz})^2+0.0035(f/1\,\mbox{kHz})^3.
$$
The angular velocity of the particles falling from the marginally stable 
Keplerian orbit along helical trajectories is at a maximum on the NS 
surface. As follows from SS 98, the formula
$$
f_s= \f{p_{*}+\omega-\omega_{*}}{r/F^2 +\omega( p_{*}+\omega-\omega_{*})}
$$
holds for the particle angular velocity near the NS surface $f_s= d\phi/dt/2\pi.$

The approximations for $f_s$ on the stellar surface and for $f_K$ in the 
marginally stable orbit as functions of the NS rotation frequency are
$$
\mbox{EOS A}\qquad f_s/1\,\mbox{kHz}\approx  2.251 -0.0236(f/1\,\mbox{kHz}) -
0.28(f/1\,\mbox{kHz})^2+0.062(f/1\,\mbox{kHz})^3; 
$$
$$
 f_K^*/1\,\mbox{kHz}\approx 1.575 +0.421(f/1\,\mbox{kHz}) -
0.065(f/1\,\mbox{kHz})^2-0.016(f/1\,\mbox{kHz})^3;
$$
$$
\mbox{EOS FPS}\qquad f_s/1\,\mbox{kHz}\approx  1.891 -0.07(f/1\,\mbox{kHz}) -
0.224(f/1\,\mbox{kHz})^2+0.194(f/1\,\mbox{kHz})^3; 
$$
$$
 f_K^*/1\,\mbox{kHz}\approx 1.561 +0.484(f/1\,\mbox{kHz}) -
0.081(f/1\,\mbox{kHz})^2+ 0.008(f/1\,\mbox{kHz})^3.
$$
We emphasize that $f_s=f_K^*$ when $R=R_*$. The softer is the equation 
of state and the higher is the NS angular velocity when it 
counterrorates with the disk, the larger is the difference  $f_s-f_K^*$.
Thus, $f_s$ specifies an additional characteristic frequency in the 
problem considered by Inogamov and Sunyaev (1999), which is appreciably 
higher than the rotation frequency of the matter in the bright 
latitudinal rings.

{\bf The NS Parameters at Which its Equatorial Radius is Equal to the Radius 
of the Marginally Stable Circular Orbit.} In order to determine the NS 
parameters at which its equatorial radius is equal to the radius of the 
marginally stable circular orbit in the equatorial plane, we must solve 
the system of equations (19) and (20) in which the radius of the 
marginally stable orbit must be substituted for $R.$ If we use formula 
(23) for the equatorial radius, which includes only one function $M(j, 
m)$ and its derivative with respect to $m$, then we obtain values that 
differ only slightly from those calculated using (19) and (20), within 
the accuracy of our calculation. In this case, however, we do not know 
the radial parameter $r = \rho^2$ to approximate the functions expressed in 
terms of the metric coefficients.

Recall that the thermodynamic function $M(j, m)$ of its own corresponds 
to each equation of state [see expressions (16) and (17) for this 
function in the case of EOS A and EOS FPS].

We use formula (47) for the radius $R_*$ in which the corresponding 
expressions [see formulas (31) and (32) for EOS FPS and EOS A, 
respectively must be substituted for the quadrupole coefficient $b = 
b(j, m).$ Thus, for each fixed $j$, we sought the corresponding solutions 
of the algebraic system (19) and (20) for $m,\, r.$  Using the 
Mathematica program, we can find a polynomial with the smallest rms 
deviation from the constructed points in the $(j, m)$  plane and obtain 
the $m =\tilde{m} (j)$ curve that separates the equilibrium positions when the 
NS lies within its marginally stable circular orbit in the equatorial 
plane from the equilibrium positions with the NS lying outside this 
orbit.

The corresponding equations of the above curve in the $ (j, \,m)$ plane are
\begin{equation}
\tilde{m}(j) =  1.228 +0.675 j - 0.159\, j^2 + 0.104\, j^3 + 2.077\, j^4 +1.25\, j^5,  
\end{equation}
for EOS A and
\begin{equation}
\tilde{m}(j) = 1.355 + 0.797\, j - 
    0.093\, j^2 - 0.343\, j^3 + 
    1.944\, j^4 + 2.86\, j^5.
\end{equation}
for EOS FPS.

For EOS FPS in the $j,\, M$  plane, the equation of the curve separating 
the states with the NS inside and outside $R_*$ is (see Fig. 8)
$$
\tilde{M}(j)\approx 1.259 + 0.575\, j - 
    0.109\, j^2 - 0.163\, j^3 + 
    0.94\, j^4 + 3.2\, j^5.
$$

{\bf Parameters of Circular Orbits on the NS Equator at $R > R_*$ for EOS A.} 
Let us introduce the parameters
$$d_m \equiv (\tilde{m}(j) - m)/(\tilde{m}(j) + 2 j m),\quad d_M
  \equiv (\tilde{M}(j) - M)/(\tilde{M}(j) + 2 j M).
$$
These parameters are zero at  $R = R_*$ and unity at $j = -0.5$. When 
$d_m$ changes from 0 to 0.5, the $d_m(j, m) =\mbox{ const}$ curves in the $(j, m)$ plane fill the region  $ D = \{  1.1\geq m \geq 2.5\}\times \{R \geq R_*\}.$ 
Accordingly, when $d_M$ changes from 0 to 0.5, the $d_M(j, M) =\mbox{ const}$ 
curves in the $ (j, M)$ plane fill the region $ D = \{  1\geq M \geq 2.1\}\times \{R \geq R_*\}.$   We therefore approximate the energy $E$, angular momentum $l$ 
[calculated using (38)], and particle angular velocity $f_K$ in a circular 
orbit lying on the stellar equator [calculated using (39)] in the region 
$D$ by polynomials in $j$ and $d_m.$ For this purpose, we first approximate 
these functions at $d_m =$ 0, 0.1, 0.2, and 0.3 by fourth-degree 
polynomials, finding the corresponding radial coordinates using Eq. 
(19), and then interpolate them in $d_m$ (or $d_M$) using 
Silvester - Legendre's formula. Again using the Mathematica program, we 
obtain for EOS A
\begin{eqnarray}
E = &0.943 - 0.029\,j + 0.019\,j^2 - 
0.017\, j^3 - 0.074\, j^4 + \nonumber\\ 
 & d_m\, (0.002 - 0.001\,j + 0.031\, j^2 +
 0.305\, j^3 + 0.416\, j^4) +\nonumber\\
 & d_m^2\, (0.097 + 0.398\, j + 0.072\, j^2 - 1.46\, j^3 -
1.56\, j^4) + \nonumber\\
 & d_m^3\, (-0.057 - 0.537\, j - 0.752\, j^2
+ 1.712\, j^3 + 3.011\, j^4)
,\\
l c/(G M) = & 3.463 - 0.909\, j + 
    0.661\, j^2 - 0.41\, j^3 - 
    1.373\, j^4 +\nonumber\\
 &  d_m\,  (0.028  + 
          0.076\, j + 1.18\, j^2 + 
          4.304\, j^3 + 3.95\, j^4) +\nonumber\\
  & d_m^2 \, (1.418 + 
          5.279\, j + 0.318\, j^2 - 
          15.653\, j^3 - 13.44\, j^4) +\nonumber\\     
 & d_m^3 \, (0.855 - 
          3.047\, j - 14.241\, j^2 + 
          8.861\, j^3 + 35.58\, j^4),\\
f_K/1\mbox{kHz}  = &  1.972 + 0.466\, j - 
1.226\, j^2 - 0.741\, j^3 + 
    2.746\, j^4  + \nonumber\\
   & d_m\, (-1.183 - 2.536\, j + 
          4.738\, j^2 - 0.986\, j^3 - 
          32.15\, j^4) + \nonumber\\
   & d_m^2\, (0.394 + 
          2.551\, j - 2.079\, j^2 + 
          22.97\, j^3 + 102.84\, j^4)  + \nonumber\\
   & d_m^3\, (-0.993 - 1.237 j + 
          13.343\, j^2 - 34.48\, j^3 - 
          169.05\, j^4).
\end{eqnarray}
 
{\bf Parameters of Circular Orbits on the NS Equator at $R > R_*$ for EOS 
FPS.} For EOS FPS, we derive the following approximation formulas for 
the parameters of an equatorial Keplerian orbit using (38) and (39):
\begin{eqnarray}
E =& \tilde{E}(j)+ d_m\,(0.01 -0.02  j -0.294\, j^2 + 
          0.196\, j^3 +1.586\,  j^4) + \nonumber\\
  & d_m^2\,(0.026 + 0.537 j +2.775\, j^2 -1.21\, j^3 - 11.039\, j^4) +\nonumber\\
  & d_m^3\,(0.103 -0.874\, j - 
         7.258\,j^2 + 1.943\, j^3 +27.516\, j^4)\approx\nonumber\\
  &\tilde{E}(j)+ d_M\,(0.002 - 0.026\, j + 
     0.043\,j^2 + 0.131\, j^3 + 
     0.373\, j^4) + \nonumber\\
    & d_M^2\,
   (0.105 + 0.748\, j - 1.184\, j^2 + 
     0.916\, j^3 + 2.726\, j^4) + \nonumber\\
    & 
  d_M^3\,(-0.069 - 1.583\, j + 5.382\, j^2 - 
     6.883\, j^3 - 14.83\, j^4),\\
&\tilde{E}(j)\approx 0.942 - 0.03 j + 0.039 j^2 -
 0.011 j^3 - 0.153 j^4;\nonumber\\ 
lc/(G M) = & \tilde{l}(j) +  
  d_m\, (0.049 - 0.147\,j + 0.237\, j^2 + 
     0.408\, j^3 + 4.79\, j^4) +  \nonumber\\
    & d_m^2\, (1.306 + 5.733\, j + 5.272\, j^2 - 
     2.106\, j^3 - 3.207\, j^4) + \nonumber\\
  &
  d_m^3\, (0.743 - 3.434\, j - 23.054\, j^2 - 
     11.233\, j^3 + 32.45\, j^4)\approx \nonumber\\
 & \tilde{l}(j)  +  
  d_M(0.041 - 0.466\, j + 2.926\, j^2 - 
     1.619\, j^3 + 1.189\, j^4) +\nonumber\\
  & 
  d_M^2(1.064 + 12.501\, j - 42.641\, j^2 + 
     63.336\, j^3 + 41.062\, j^4) +\nonumber\\
  & 
  d_M^3\,(2.512 - 25.467\, j + 182.594\, j^2 - 
     284.35\, j^3 - 202.93\, j^4),\\
 &\tilde{l}(j)\approx 3.457  -0.924\, j +0.895\, j^2 - 
  0.375\, j^3 -  2.224\, j^4;\nonumber\\
\Omega_K G M/c^3 = & \tilde{f}(j) + 
  d_m(0.083 + 0.184\, j - 0.013\, j^2 - 
     0.087\, j^3 - 0.116\, j^4) + \nonumber\\ 
&  d_m^2\,(-1.013 - 2.31 j - 0.268 j^2 + 
     1.116\, j^3 + 2.894\, j^4)+\nonumber\\
  & 
  d_m^3\,(1.794 + 4.686\, j + 3.981\, j^2 - 
     2.428\, j^3 - 17.387\, j^4)\approx\nonumber\\
  &  \tilde{f}(j) + 
  d_M\,(-0.097 - 0.17 j + 0.071 j^2 - 
     1.156\, j^3 - 2.302\, j^4) + \nonumber\\
  & 
  d_M^2\,(-0.136 - 0.754\, j + 1.681\, j^2 + 
     11.715\, j^3 + 9.366\, j^4)+\nonumber\\
  & 
  d_M^3\,(0.614 + 3.075\, j - 5.577\, j^2 - 
     32.591\, j^3 - 13.9\, j^4),\\
& \tilde{f}(j)\approx 0.059 + 0.025\, j - 0.062\, j^2 - 
  0.017\, j^3 + 0.148\, j^4.\nonumber
\end{eqnarray}

Here, $\Omega_K=2\pi f_K$ is the particle angular velocity in the equatorial Keplerian 
orbit (in rad/ s).

{\bf The Maximum Rotation Frequency in Circular Orbits at $R \leq
R_*$ and  $R > R_*$.} The marginally stable orbits [see formula (48)] and the Keplerian 
orbits lying on the NS surface [see formulas (53) and (56)] have the 
largest frequency in Keplerian equatorial orbits around the NS at  $R \leq
R_*$ and $R > R_*$, respectively. We emphasize that we deal with circular 
orbits; for the spiraling-in particles in the gap between $R_*$ and $R$, 
the angular velocities $f_s$ are higher than those in the marginally 
stable orbit (see above). In order to interpret the quasi-periodic 
millisecond oscillations from LMXB objects, it is useful to calculate 
the dependence of the largest Keplerian frequency on the NS angular 
velocity; either (48) or (53) and (56) must be used, depending on the 
situation. In Fig. 10, maximum Keplerian frequency is plotted against NS 
rotation frequency for EOS FPS for fixed rest masses $m$ at 0.1 steps. 
The NS angular velocity was calculated by using (9). The corresponding 
curves are indicated by solid and dashed lines at $R \leq
R_*$ and $R > R_*$, 
respectively. The curve for $R = R_*$ is also indicated by a dashed 
line. When this curve is reached, the maximum possible Keplerian 
frequencies are obtained for a fixed rest mass. The dots indicate the 
curve of stability loss according to the static criterion; it correspond 
to the $M = M_*(j)$ curve [$m = m_*(j)$] [see formulas (14) and (16)].
\begin{figure}[t]
\includegraphics[width=0.75\linewidth,bb=94 230 560 724,clip]{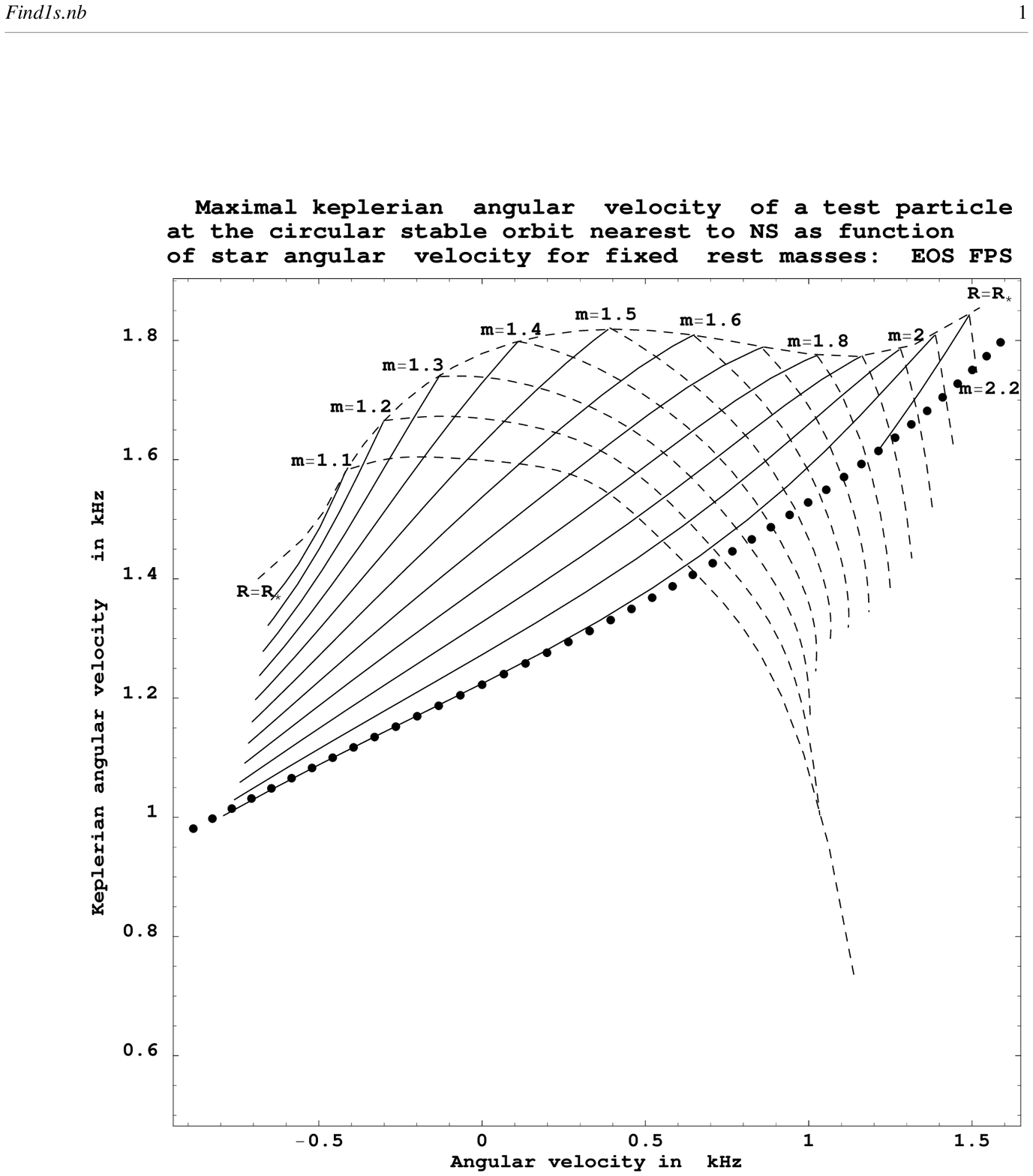}
\caption{\small  Particle rotation frequency in the marginally stable circular
Keplerian orbit at $R < R_*$ (solid lines) and particle rotation frequency
in an equatorial Keplerian orbit at $R > R_*$ (dashed lines) versus NS
rotation frequency for fixed rest masses $m$ (EOS FPS). The upper envelope
specifies the frequency for $R = R_*;$ at $m =$ const, the functions
$f^*_K(f)$ monotonically increase for $R < R_*,$ while the functions $f_K(f)$
monotonically decrease for $R > R_*.$}
\end{figure}
\section{ NS spinup via accretion and its luminosity.}

{\bf Spinup in the Newtonian Theory.} Let us first consider the Newtonian 
pattern for an incompressible, self-gravitating rotating mass of fluid 
whose equilibrium figure is a MacLaurin spheroid.

The spheroid eccentricity $e$ is defined as $e=\sqrt{1-c^2/a^2}$ [$a,\,c$
are the bigger and smaller semiaxes of the spheroid respectively]. 

 The Keplerian 
equatorial rotation frequency can be determined by equating the gravity 
and the centrifugal force:
\begin{equation} 
 f_K=\sqrt{ G\rho/2\pi}B(e);\quad     B(e)\equiv\sqrt{\f{
\sqrt{1-e^2}}{e^3}\arcsin   e   -\f{1-e^2}{e^2}}   
\end{equation}
The formula relating the angular frequency and the eccentricity of a 
Maclaurin spheroid follows, in particular, from equilibrium conditions:  
 \begin{equation}  f = \sqrt{ G\rho  g(e)/ 2\pi};\quad  g(e)\equiv  \f{
\sqrt{1-e^2}}{e^3}(3-2 e^2)\arcsin e -3\f{1-e^2}{e^2}. 
\end{equation} 
Denote the mass brought by the accreting particles from the disk to the 
stellar equator per unit time, which subsequently spreads over the star, 
changing its mass, angular momentum, and total energy in an equilibrium 
way, by $\dot  M$. From the law of conservation of angular momentum, we have
\begin{equation} \f{d}{dt}I f =
\dot{ M} a f_K. \end{equation}
We derive the following evolutionary equation for the eccentricity from 
formula (59) (see also Finn and Shapiro 1991):
\begin{equation}
\f{d\,e}{d\,t}
=
\f{5(B(e)-\f23\sqrt{g(e)})}{2(2 e\sqrt{g(e)}/(3(1-e^2))+\f{d}{de}\sqrt{g(e)})}\dot{M}/M.
\end{equation}
When the disk and the Maclaurin spheroid counterrotate, the sign in Eq. 
(53) before $B(e)$ must be changed to the opposite one.

Let us introduce a dimensionless parameter of angular acceleration $q$:
\begin{equation} q \dot{M}/M = \f{d f}{d
 t}\sqrt{\f{2\pi}{G\rho}} =\f{d\,\sqrt{g(e)}}{d\,e}\f{d\,e}{d\,t}.\end{equation}

We substitute the right-hand part of (60) for $de/dt.$

The right-hand part of (61) can be expanded in a Taylor series in $e$ by 
using (58). Passing to a Taylor expansion of $q$ as a function of 
dimensionless angular velocity $\tilde{f}\equiv f\sqrt{2\pi/G\rho}$  or the ratio of the NS rotation 
frequency to the rotation frequency in an equatorial Keplerian orbit, we 
obtain
 $$
q\approx 2.0412- 5/3 \tilde{f}-5.868 \tilde{f}^2 + 25/6
\tilde{f}^3 - 0.371\tilde{f}^4 + 0.744\tilde{f}^5\approx
2.041-1.361\f{f}{f_K} \,-
$$
\begin{equation}
3.912\left(\f{f}{f_K}\right)^2+2.608\left(\f{f}{f_K}\right)^3+1.791\left(\f{f}{f_K}\right)^4
- 1.194\left(\f{f}{f_K}\right)^5.\tag{$61a$}
\end{equation}
Note that formulas (61a) are also valid for negative $f$ 
(counterrotation). The parameter of NS angular 
acceleration is determined solely by its density and angular velocity.

{\bf Spinup in General Relativity.} In general relativity, the dimensionless 
parameter of angular acceleration $q$ can be introduced as follows:
$$
\f{df}{dt}=\f{c^3}{G M_{\odot}}\f{\dot{M}}{M} q \quad \mbox{or}\quad\f{df/1
\mbox{kHz}}{dt}=111.37\f{\dot{M}}{M}q.
$$
Using our results from SS 00, we can easily obtain for $q$ in general 
relativity
\begin{equation}
q/M_{\odot} =\f{ lc/(G ) - 2 j M M_{,m}}{M +2 j M_{,j}}\left( \f
{M_{,j}}{M(M + 2 jM_{,j})}\right)_{,j} +\left( \f
{M_{,j}}{M(M + 2 jM_{,j})}\right)_{,m} M.
\end{equation}

For slowly rotating stars ($j <0.1$), the angular acceleration [as follows 
from (62)] is given by
\begin{equation}
q\approx\f{M_{\odot} M_{,jj}}{M^2}.
\end{equation}
At $M = 1.4\, M_{\odot}$, $q \approx$ 9.55, 8.33, and 8.54 for EOS A, EOS AU, and EOS FPS, 
respectively. We emphasize that, here, we consider the instantaneous 
rate of change in the rotation frequency of a NS with a given 
gravitational mass and angular velocity rather than the evolution of NS 
parameters during accretion.

Below, we restrict ourselves to numerically constructed dependences for 
the EOS FPS alone.

The approximation formula for $q$ in the case of NS and disk corotation 
at $M = 1.4\, M_{\odot}$ is
\begin{equation}
 q  \approx  10.964 - 6.057(f/1\mbox{kHz}) + 1.54\,( f/1\mbox{kHz})^2.
\end{equation}

This formula is valid for angular velocities from 100 to 800 Hz. In the 
case of accretion-disk and NS counterrotation (when the NS spins down), 
the approximation formula for $|q|$ in the same range of angular 
velocities is
\begin{equation}
 q\approx 10.565 -2.4 (|f|/1\,\mbox{kHz})  
   -4.34\,( f/1\,\mbox{kHz})^2.
\end{equation}
For $M=1.8\, M_{\odot}$, an approximate formula for the parameter of angular 
acceleration $q$ is
\begin{equation}
 q \approx  22.234 -9.664 (f/1\,\mbox{kHz})  - 1.63\, (f/1\,\mbox{kHz})^2
\end{equation}
in the case of corotation and
\begin{equation}
q
\approx 20.27 -2.96 (|f|/1\,\mbox{kHz})- 2.59\,(|f|/1\,\mbox{kHz})^2
\end{equation}
in the case of counterrotation. These formulas are valid in the range of 
angular velocities from 150 to 1200 Hz.
\begin{figure}[t]
\includegraphics[width=0.75\linewidth,bb=110 240 550 720,clip]{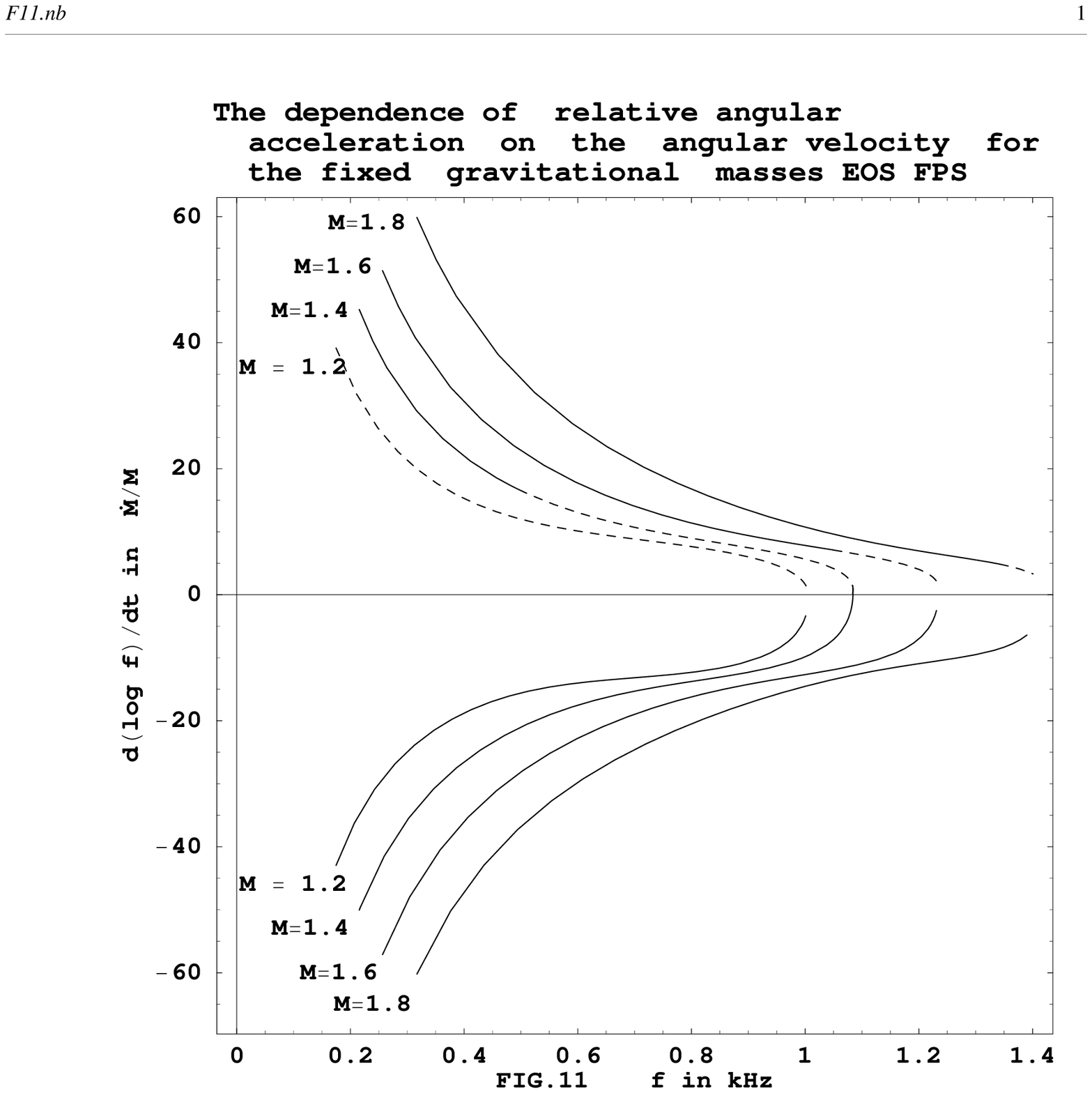}
\caption{\small  Angular acceleration via disk accretion versus NS rotation 
frequency at $R < R_*$ (solid lines) and at $R > R_*$ (dashed lines) for
fixed gravitational masses $M$ (EOS FPS) (see the approximation formulas 
(54 - 58) for $q(f)$ at the fixed gravitational masses $M = 1.4$ and 
$1.8 M_{\odot}$). The lower curves describes counterrotation (spindown). The 
upper curves describe spinup when the NS and the disk corotate. }
\end{figure}
Formulas (64) - (67) are slightly inaccurate for slow rotation. In this 
case, formula (63) must be used. The transition from corotation to 
counterrotation of the star and the disk in the exact statement 
(61)--(63) is smooth. In Fig. 11, $ 1\,\mbox{kHz}\,\,q/\,f$ is plotted against $f$ for 
fixed gravitational masses $M =1.2,\,1.4,\,1.6,\,1.8\,M_{\odot}$.

 For 
comparison, Fig. 12 shows the  $(M/\dot{M}) d\ln{f}/dt$ dependences in the Newtonian 
theory and in general relativity. The density ( $\sim  10^{14} g/cm^3$) was 
chosen in such a way that the maximum possible angular  velocity were 
equal in both approaches.
\begin{figure}[t]
\includegraphics[width=0.75\linewidth,bb=91 259 540 717,clip]{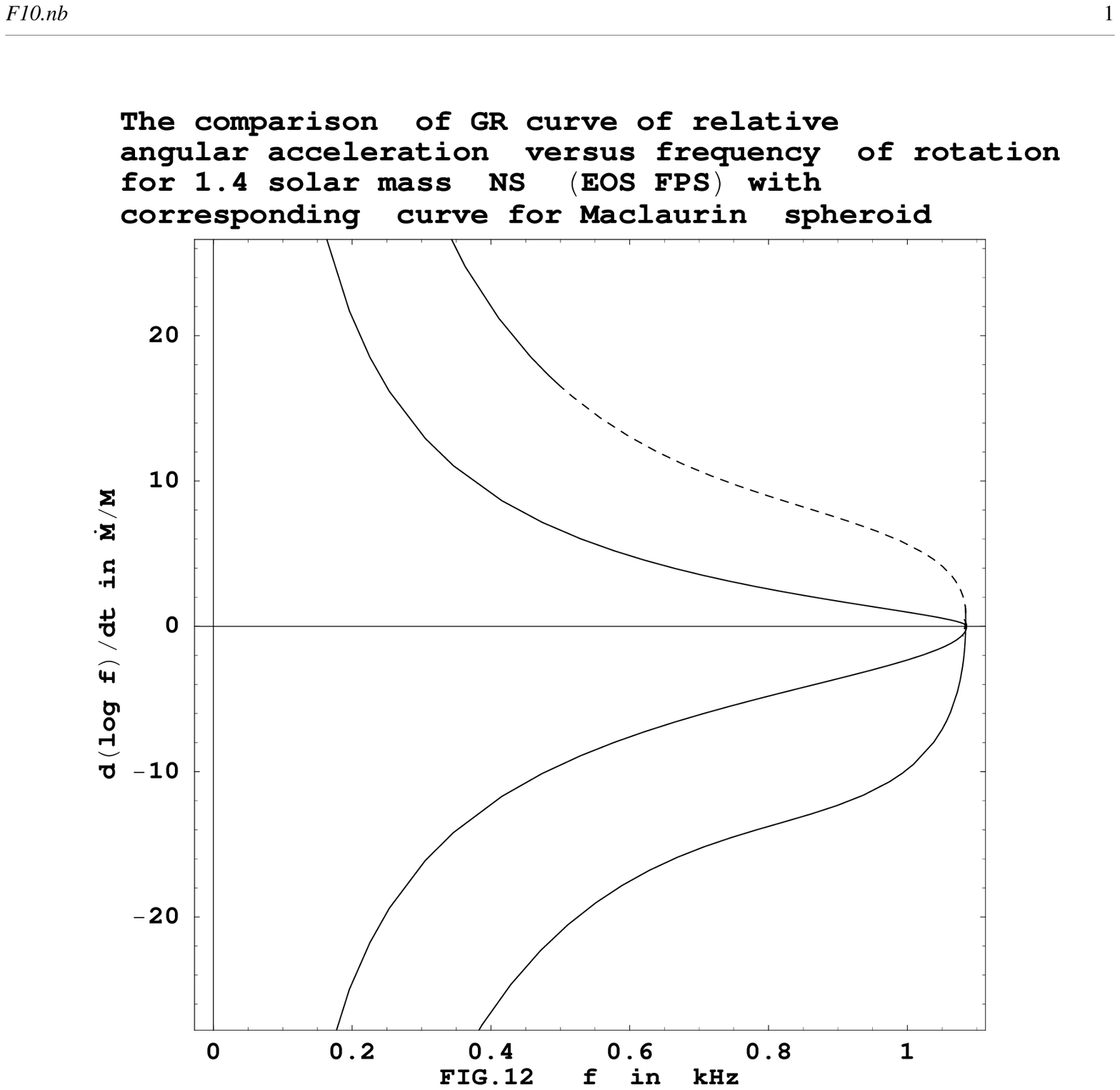}
\caption{\small  Comparison of the plots of angular acceleration $q$ versus NS 
rotation frequency for the fixed gravitational mass $M = 1.4 M_{\odot}$ in 
general relativity (EOS FPS) (heavy line) and in the Newtonian theory 
(thin line) at matched maximum angular velocities. }
\end{figure}
{\bf Energy Release on the NS Surface During Disk Accretion for Weak 
Magnetic Fields.} In Figs. 3 and 4, total energy release on the surface 
of a NS with EOS FPS and in the accretion disk (solid line), as well as 
the ratio of these energy releases (dashed line), are plotted against NS 
angular velocity for normal sequences with the fixed rest mass   $m= 1.56\, M_{\odot}$  ($
 M=1.4\, M_{\odot} \mbox{in the static limit })$ and $m =2.1\, M_{\odot}$ ($\,M = 1.8\, M_{\odot} \mbox {in the static limit })$.

We derived formulas for the energy releases in SS 00:
\begin{equation} L_s= \dot{M} c^2( E - \Omega l/c^2 - \mu),\quad L_d= \dot{M} c^2(1-
E)\quad \mbox{ at}\,R\geq R_*; 
\end{equation}
\begin{equation}
 L_s= \dot{M} c^2( E_* - \Omega l_*/c^2 - \mu),\quad L_d= \dot{M} c^2(1-
E_*)\quad \mbox{at}\,R\leq R_*.
\end{equation}
Here, $\mu$ is the NS chemical potential, and $\Omega$ is its angular 
velocity (see SS 00). The quantities in the formulas for the surface 
energy release are given by the following formulas: (9) for $\Omega$; 
(19) for $\mu$; (45), (46) for $E_*, l_*$; (51), (52) for EOS A and (54), 
(55) for EOS FPS for $E, l.$ In the case of a nonrotating NS, the 
formulas for gravitational energy release were derived by Sunyaev and 
Shakura (1986).

The energy release reaches a maximum when the disk and the neutron star 
counterrotate: for EOS FPS, it is $0.414 \dot{M}c^2$ at $f = -1.08\,$ kHz
{\t for fixed rest mass} $m =1.56\, M_{\odot}\,$ and $\,0.67 \dot{M}\,c^2$
at  $f = -1.49\,$ kHz for fixed $\,m =2.1\, M_{\odot}$.

Figures 1 and 2 show these quantities for the {\it fixed gravitational masses} 
$\,M = 1.4\, M_{\odot} $ and $M = 1.8\, M_{\odot}$.

The maximum luminosity for the gravitational mass  $M
 =1.4 M_{\odot}$ is  $0.408 \dot{M}\,c^2$ 
 at $f = -1.08$ kHz. The approximations for the total 
luminosity and the $L_s/(L_s+L_d)$  ratio at $|f| < 1$ kHz for the 
gravitational mass $M
 =1.4 M_{\odot}$  are  given by formulas (1,2) in the
Introduction.

The maximum luminosity for $M
 =1.8\, M_{\odot}$   is  $0.62
 \dot{M}c^2$ at $f = -1.41$ 
kHz. The approximation for the total luminosity at $|f| < 1 kHz$ for $M
 =1.8\, M_{\odot}$  is
\begin{equation}
L_s + L_d\approx (0.333 - 0.193\, f/1\,\mbox{kHz}+ 0.006\, (
f/1\,\mbox{kHz})^2) \dot{M} c^2.
\end{equation}
If the $f$ dependence of $L_s/(L_s + L_d)$ is approximated by a quadratic 
trinomial in the range $ -1\,\mbox{ kHz} < f < 1 \,\mbox{kHz}$ at $M
 =1.8\, M_{\odot}$, then we obtain
\begin{equation}
L_s/(L_s + L_d) \approx 0.83 - 0.17\, f/1\,\mbox{kHz} - 0.1\,
(f/1\,\mbox{kHz})^2.
\end{equation}
For comparison, we give approximations for $\epsilon_{grav}=(L_s+L_d)/\dot{M}$, the 
efficiency of gravitational energy release on the surface of a NS with 
the soft EOS A and the moderate EOS AU, as functions of rotation 
frequency $f$ for  NS normal sequence with $M= 1.4 M_{\odot}$ in the static limit:
\begin{equation}
\mbox{EOS A}:\qquad\epsilon_{grav}/c^2=0.245-0.152\, f/1\,\mbox{kHz}+0.02\,(f/1\,\mbox{kHz})^2,
\end{equation}
\begin{equation}
L_s/(L_s+L_d)=0.765-0.242\, f/1\,\mbox{ËçÃ}-0.108\,(f/1\,\mbox{kHz})^2,\qquad
|f|<1.3 \mbox{kHz};
\end{equation}
\begin{equation}
\mbox{EOS AU}:\qquad\epsilon_{grav}/c^2 = 0.225 - 0.137\, f/1\,\mbox{kHz}+0.01\,(f/1\,\mbox{kHz})^2,
\end{equation}
\begin{equation}
L_s/(L_s+L_d)=0.773 - 0.301\, f/1\,\mbox{kHz}-0.174\,(f/1\,\mbox{kHz})^2,\qquad
|f|<1 \mbox{kHz}.
\end{equation}
The softer is the equation of state, the stronger is the concentration 
of matter toward the stellar center, and the larger is the gap between 
the marginally stable orbit and the NS surface. Therefore, the energy 
release on the surfaces of stars with a soft equation of state at the 
same masses and angular velocities exceeds the energy release on the 
surfaces of stars with a stiff equation of state.

As follows from (1), (70), (72), and (74), the total energy release is a 
nearly linear function of the NS rotation frequency over a wide range of 
its variation, $|f| < 1 kHz.$ This distinguishes the general-relativity 
results from the Newtonian theory of disk accretion onto Maclaurin 
spheroids (see below).

{\bf Estimating the Energy Release in the Newtonian Theory.} When a rotating 
NS is modeled by a Maclaurin spheroid, the following formula holds for 
the total energy release in the disk and on the NS surface ($\phi_e$ is 
the gravitational potential at the equator):
\begin{equation}
\f{L_s+L_d}{\dot{M} R^2}= \f{1}{2}  ((\Omega_K-\Omega)^2+2\f{\phi_e}{R^2}-\Omega_K^2)=\pi
G\rho((B-\sqrt{g})^2+2\f{(1-e^2)^{3/2}}{e^3}\arcsin{e}+2-2e^2).
\end{equation}
The notation (57) and (58) is used in (76). Expanding $B(e)$ and $ g(e)$ in 
a Taylor series in powers of $e$ and expressing $e$ in terms of  $\tilde{f}\equiv f\sqrt{2\pi/G\rho}$ and $f/f_K$ yield
$$
\f{L_s+L_d}{0.11\dot{M} c^2}\left(\f{M}{1.4
M_{\odot}}\right)^{-2/3}\left(\f{\rho}{10^{14}\mbox{g/cm}^3}\right)^{-1/3}\approx 1- 1.2247
\tilde{f}+0.5\tilde{f}^2-1.0716\tilde{f}^3+0.3125\tilde{f}^4 - 1.608\tilde{f}^5\approx
$$
\begin{equation}
1-\f{f}{f_K}+\f{1}{3}\left(\f{f}{f_K}\right)^2-\f{1}{3}\left(\f{f}{f_K}\right)^3-\f{1}{36}\left(\f{f}{f_K}\right)^4+\f{1}{36}\left(\f{f}{f_K}\right)^5.
\end{equation}
The speed of light in (77) was introduced for convenience of comparing 
the energy releases in the Newtonian theory and in general relativity.

For the ratio of the energy release on the surface of a Maclaurin 
spheroid and the total energy release, we have an exact formula:
\begin{equation}
\f{L_s}{(L_d+L_s)}=0.5(1-f/f_K)\approx 0.5-0.612\tilde{f}-0.23\tilde{f}^3-0.5\tilde{f}^5-1.33 \tilde{f}^7. 
\end{equation}

Note that, in contrast to general relativity, the Newtonian theory 
underestimates the contribution of the surface luminosity to the total 
luminosity for slow rotation and counterrotation: compare formulas (78) 
with (2), (71), (73), and (75).

\section{ Neutron Star and Disk Counterrotation }

The approach developed here allows the energy release during disk 
accretion to be calculated in the two most important cases where the 
star and the disk matter corotate and counterrotate. Unfortunately, the 
problem with an arbitrary angle between the disk and NS rotation axes is 
much more complex.

As we see from the figures and the approximation formulas (1), (2), 
(70 - 75), when the directions of the rotation axes coincide, the 
surface energy release decreases in importance compared to the static 
case, while the disk energy release increases in importance. An 
important thing is that the gap between the disk and the NS disappears 
for rapid rotation in the same sense. The total energy release is 
appreciably smaller (by a factor of 1.55) than that for a nonrotating NS 
with the external geometry described by the Schwarzschild solution even 
at the observed rotation periods of bursters (of the order of 600 Hz). 
The $L_d/L_s$ ratio is equal to unity at $f = 600 Hz$ for EOS FPS and $M=1.4 M_{\odot}$.

Generally, counterrotation is great interest. In this case, the total 
energy release during disk accretion abruptly increases, reaching the 
record $0.49 \dot{M}c^2$ for an $1.4 M_{\odot}$ star in the case of the most rapid 
counterrotation, and even reaches $0.67 \dot{M}c^2$ for the maximum-mass normal 
sequence, which is unstable in the static limit for a NS with the soft 
EOS A. Note that this energy release exceeds appreciably the energy 
release of 0.422$ \dot{M}c^2$ in the disk during accretion onto a Kerr black hole 
with the maximum possible dimensionless angular momentum $j = 1.$ 
Interestingly, by contrast to a Kerr black hole, the energy release 
during accretion onto a NS is at a maximum in the case of 
counterrotation. In section 5, we provide approximation formulas for the 
other two equations of state as well, showing that there is the same 
tendency for them. In general relativity, enhanced energy release is 
accompanied by a faster angular deceleration of the NS than in the 
Newtonian theory. As we see from Figs. 1-4 and the approximation 
formulas (24 - 28), there is a gap between the marginally stable orbit 
and the NS surface in the case of counterrotation almost for all 
equations of state.

We do not consider the detailed physics of the spread and boundary 
layers but discuss only the parameters of the marginally stable orbit 
disregarding other forces acting on the accreting particles, for 
example, light-pressure forces. Concurrently, the $L_s/L_d$ ratio abruptly 
increases for counterrotation: much more energy is released on the NS 
surface than in the disk. Formulas (2), (71), (73), and (75) give simple 
approximations of $L_s/(L_s+L_d)$ for three characteristic NS equations of 
state as functions of stellar rotation frequency.

{\bf Nuclear and Gravitational Energy Release in X-ray Bursters.} The most 
important parameter of X-ray bursters is the ratio of energy $E_{nuc}$ 
released during a relatively short (5-20 s) X-ray burst (resulted from 
a nuclear explosion in the matter accreted onto the NS surface) to 
energy $E_{grav}$ released via accretion in intervals between two successive 
explosions $T.$ This energy release of the infalling matter is related to 
the release of gravitational energy. Explosions recur as nuclear fuel is 
accumulated and follow with a quasi-period from several hours to several 
days. Clearly,
 $$
\f{E_{nucl}}{E_{grav}}=\f{\int_{t_{burst}}
L_{burst}\,dt}{\int_{t_i}^{t_i+T}L_{grav}\,dt}=\f{\epsilon_{nucl}}{\epsilon_{grav}},
$$
where $\epsilon_{nucl}$ is the efficiency of nuclear energy release, close to 1 
MeV/nucleon  $\approx10^{-3} c^2,$ for helium burning and its conversion into  $C^{12}$ 
and for the subsequent thermonuclear reactions up to the production of 
iron (Bildsten 2000). As we pointed out above, $\epsilon_{grav}=(L_s+L_d)/\dot{M}$
is a strong (perceptible) function of the NS angular velocity and its direction.

For a 1.4 $M_{\odot}$ star, the energy release $\epsilon_{grav}$ during accretion onto a NS is 
(depending on the equation of state): (1) 0.245$ c^2$ for EOS A, 0.217$ c^2$ 
for EOS AU, and 0.213$ c^2$ for EOS FPS in the case of a slowly rotating 
star; (2) 0.172$ c^2$ for EOS A, 0.145$ c^2$ for EOS AU, and 0.128$ c^2$ for EOS 
FPS in the case of corotation with a frequency of 600 Hz; and (3) 
0.32$ c^2$ for EOS A, 0.311$ c^2$ for EOS AU, and 0.308$ c^2$ for EOS FPS in the 
case of counterrotation with a frequency of 600 Hz.

It is thus clear that allowance for rotation can account for the 
observed luminosity variations from source to source within a factor of 
2 - 2.5.

In this case, only $\epsilon_{grav}$ varies; we assume $\epsilon_{nucl}$ to be independent of 
the stellar rotation and ignore the difference in the spectra of the 
disk and the stellar surface.

This simple argument indicates that $\epsilon_{grav}/\epsilon_{nucl}$  is relatively low for 
corotating objects. At the same time, the considerably rarer cases of 
counterrotation must result in abnormally high $\epsilon_{grav}/\epsilon_{nuc}$; this cannot 
affect the recurrence time of nuclear bursts, but appreciably reduces 
their amplitude. It is reflected only in an increase of the persistent 
flux between bursts. Such cases are observed. One should pay particular 
attention to such sources as objects in which the angle between the NS 
and disk rotation axes exceeds appreciably $\pi/2.$

{\bf Observational Differences between Corotating and Counterrotating 
Objects.} The observational manifestations of accreting neutron stars 
must strongly depend on whether the NS and the disk corotate or 
counterrotate even at observed rotation frequencies of $\sim$600 Hz. The most 
important difference is associated with the presence of a fairly 
extended gap between the disk and the NS when they counterrotate. This 
may give rise to an appreciable energy release in the equatorial 
boundary region, in addition to the two regions equidistant from the 
equator predicted by the theory of matter spread over the NS surface. At 
low accretion rates, a hard spectrum originating in a strong shock wave, 
in which the radial velocity is lost, can form in the equatorial region. 
Thus, the radial and azimuthal velocities determine the energy release 
in the equatorial region and in the bright equidistant belts (Inogamov 
and Sunyaev 1999), respectively. At all the observed NS rotation 
frequencies, the energy release associated with the radial velocity 
(though it reaches 10 - 15 \% of the azimuthal velocity in the equatorial 
region on the NS surface) is incapable of significantly affecting the dynamics of the infalling gas, because it cannot compensate for the 
difference between the gravity and the centrifugal force for the 
infalling particles near the NS surface. The presence of a gap at low 
accretion rates allows us to record photons from the lower NS 
hemisphere, which is completely hidden by the disk from the observer in 
the case of corotation where there is no gap. However, the main thing is 
that the surface luminosity greatly exceeds the luminosity in the 
accretion disk when the $L_s/L_d$ ratio reaches 5 or 6 for counterrotation 
of an  $M = 1.4\,M_{\odot}$ star with a frequency of 600 Hz. Note that, when the star cororates and has a rotation frequency of 600 Hz, the disk 
luminosity is approximately the same as that of the entire NS surface. 
This case is much closer to the case of accretion onto a black hole 
(where there is no solid surface at all) than the case of 
counterrotation.

{\bf Why Might We Expect the Existence of Accretion Disks around 
Counterrotating Stars?} In the standard pattern of NS spinup by the 
accreting matter, all stars must eventually corotate with the accretion 
disk. Nevertheless, nature allows for the formation of counterrotating 
low-mass star and disk.

(1) Imagine a binary produced by an explosion of the more massive 
component turning into a NS. The kick during an asymmetric explosion, 
which is widely discussed in the literature, must result not only in a 
high velocity of the formed NS but also in its appreciable rotation 
(Spruit and Phinney 1998). In this case, it is doubtful that the sense 
of rotation resulting from the kick coincides with the sense of rotation 
of the disk, which is determined by the direction of the binary's 
orbital angular momentum. The circularization time of a close binary's 
orbit must be appreciably shorter than the time of change in the NS rotation axis.

(2) Tens of millisecond pulsars are observed in rich globular clusters, 
and there may be many hundreds of binaries containing neutron stars with 
weak magnetic fields. A change of the low-mass partner in such a binary 
during a close encounter with a single star (see McMillan and Hut 1996) 
can, in principle, give rise to a binary with NS and disk 
counterrotation. The ejection of such a binary from the globular cluster 
by tidal forces when the cluster traverses the central part of the 
Galaxy (or through the interaction of close pairs inside the globular 
cluster) can give rise to binaries with NS and disk counterrotation 
outside these clusters.

(3) If a rapidly rotating neutron star with a weak magnetic field is 
born during gravitational collapse in a massive binary, then accretion 
of the high-speed stellar wind emitted by a massive, hot supergiant takes place. Illarionov and Sunyaev (1974) pointed out that, in this 
case, an accretion disk can form around the NS, but the sense of 
rotation of the disk can depend on density and velocity fluctuations of 
the matter near the capture radius. Clearly, the transition from states 
close to NS and disk corotation to states with an appreciable angle 
between the NS and disk rotation axes must be accompanied by a radical 
change in the effective energy release and spectrum. If a 
counterrotating disk is formed in this case, then surface nuclear 
explosions are much more difficult to observe in such a binary because 
of the low $\epsilon_{nucl}/\epsilon_{grav}$ ratio, which may hamper the binary's identification as an X-ray burster. Such a binary may turn out to be 
similar in many ways to 4U 1700--37, from which no regular X-ray 
pulsations are observed (there is no strong magnetic field) and from 
which (although the X-ray emission is erratic) no X-ray bursts of the 
first type (coupled with surface nuclear explosions) have been observed 
so far.

\section{Discussion} 

Rapid rotation of accreting neutron stars is widely discussed in the 
literature devoted to interpretation of the nature of kilohertz 
quasi-periodic X-ray oscillations from low-mass X-ray binaries (Van der 
Klis 2000; Wijnands and Van der Klis 1997; Miller et al. 1998; 
Stromayer et al. 1998; Titarchuk and Osherovich 1998). The kilohertz 
quasi-periodic oscillations were discovered from the RXTE satellite.

For neutron stars rotating with periods of 300--600 Hz, the NS rotation 
affects significantly its internal structure and external field. Hartle 
and Sharp (1967) (see also Hartle 1978) developed a variational 
principle for rotating barotropic stars.

Numerical calculation of a rapidly rotating star with a polytropic 
equation of state, calculation of differential rotation, and the case of 
piecewise constant polytropic indices (as the asymptotics of the 
equations of state for stellar matter at temperatures below the 
degeneracy temperature; see Eriguchi and Mueller 1985) for large angular 
velocities present series computational difficulties even in the 
Newtonian approximation. These difficulties have been overcome only 
relatively recently (see Ostriker and Mark 1968; Tassoul 1978; Hachisu 
1986). In general relativity, there are several numerical algorithms 
(codes) for finding steady-state configurations of rotating gas masses 
in their gravitational fields: Butterworth--Ipser's (1976) method, 
Friedman et al.'s (1986) modification of this method, the KEH method 
(Komatsu et al. 1989), Cook et al.'s (1994) modification of this 
method (see also Stergioulas and Friedman 1995), and the BGSM code based 
on spectral methods (Bonazzola et al. 1993;  see the comparison
of different approaches  Eriguchi et al. 1994; 
Nozawa et al. 1998). Friedman et al. (1986) first published 
calculations of steady-state configurations of rapidly rotating neutron stars using realistic tabulated equations of state. Previously, Butterworth and 
Ipser (1976) and Bonazzola and Schneider (1974) used polytropic models 
and an incompressible fluid model. Many important physical parameters of 
the NS external field for 14 equations of state (including detailed 
tables of sequences with a fixed rest mass for EOS A, AU, FPS, M, and L) 
are contained in Cook et al. (1994). Based on the BGSM code, Salgado 
et al. (1994) analyzed the global parameters of neutron stars with 
different equations of state.

Datta et al. (1998) gave tables of NS parameters at fixed NS angular 
velocity and central density for EOS A, B, C, D, E, F, and the new 
equations of state BBB1, BBB2, BPAL21, and BPAL32. Based on the KEH 
method, Stergioulas (1998) developed a numerical code for computing the 
global parameters of neutron stars with 
many known equations of state. 
This code computes the NS properties as functions of two parameters: one 
is the central density, and the other can be either the rest mass or the 
gravitational mass, or the angular momentum, or the angular velocity. We 
repeatedly used this code in our calculations, which allowed us to 
construct the above approximation formulas.

As was already pointed out above, the gravitational fields of rotating 
gas configurations are essentially nonspherical at large angular 
velocities (Chandrasekhar 1986); all multipole components [as, for 
example, on the surface of a Maclaurin spheroid; see formulas 
(76) - (78)] contribute to the gravitational field near them.

Approximate and exact analytic solutions of the Einstein equations in a 
vacuum which approximate the numerical results obtained by the above 
authors are of considerable interest in describing the physical 
processes in strong external gravitational fields.

This problem was independently considered by several authors in 1998. 
Based on the numerical results of Cook et al.(1994), Laarakkers and 
Poisson (1998) analyzed the dependence of the quadrupole coefficient 
[which emerges in the expansion of the metric potential $\nu$ in the 
form of Bardeen and Wagoner (1971) at large radii] on Kerr parameter $j.$ 
Laarakkers and Poisson (1998) pointed out that, in contrast to the 
nonrelativistic case, this dependence can be approximated by a parabolic 
law for the equations of state they considered.

Shibata and Sasaki (1998) used Fodor et al.'s (1989) asymptotic 
expansions of the function $\xi = \sqrt{\rho^2+z^2}(1-{\bf E})/(1+{\bf E})$
(here, ${\bf E}$ is the Ernst complex 
potential; $\rho$ and $z$ are Weil's canonic coordinates) in the 
equatorial plane and on the symmetry axis to determine the radius of the 
marginally stable orbit and the particle angular velocity in this orbit 
in the form of formal expansions in powers of the dimensionless angular 
momentum $j.$ The coefficients of these expansions were expressed in 
terms of Geroch--Hansen's multipole coefficients (see Hansen 1974), 
relative to which their order by $j$ was assumed.

The marginally stable orbit is of considerable importance in 
interpreting the kilohertz QPOs detected by the RXTE satellite. 
Therefore, based on the numerical results of Cook et al. (1998), 
Miller et al. (1998) analyzed the dependences of the radius and 
angular velocity in the marginally stable orbit, as well as the 
equatorial radius, on the NS angular velocity for various equations of 
state of neutron matter.

Independently, Thampan and Datta (1998) also numerically analyzed the 
dependence of the Keplerian angular velocity in the marginally stable 
orbit on the NS angular velocity. When calculating the luminosity from 
the equatorial boundary layer, these authors assumed that the accreting 
particles radiated away all the energy equal to the difference between 
the energy in a Keplerian orbit and the energy of the particle 
corotating with the star on its equator. This assumption differs 
radically from our treatment in SS 98, SS 00, and this paper.

When the inverse effect of the spreading matter is considered on long 
time scales, the structural changes in the neutron star caused by the 
changes in its mass and angular momentum through their influx during 
accretion must be take into account. Lipunov and Postnov (1984) and 
Kluzniak and Wagoner (1985)  in calculation of NS spin-up by the
disc accretion assumed the moment of inertia to be equal to that of
nonrotating star 
and the rotation velocities to be low enough.

Burderi et al. (1998) made an attempt to describe the evolution of the 
NS angular velocity under the effect of disk accretion by using the Kerr 
metric for the NS external field. These authors also used approximation 
formulas for the gravitational mass of the type $ M =m(1-\alpha/R),\qquad M  R^3=\mbox{const}$
 (here, $m$ is the NS rest mass) and extrapolated the formula of 
Ravenhall and Pethick (1994) for the moment of inertia in the static 
case to the case of rapid rotation; they postulated a relationship 
between the equatorial radius and the angular velocity, expressions for the maximum radius and the maximum angular velocity, etc.

Our approach (SS 00 and this paper) assumes that the luminosity from the 
stellar surface is considered on the basis of the first law of 
thermodynamics by taking into account changes in the NS total energy 
during quasi-uniform changes in its parameters under the effect of disk 
accretion. In the Newtonian problem, the energy release is proportional 
to the square of the Keplerian velocity on the NS equator relative to 
the frame of reference corotating with the star. This idea was 
generalized to the case of general relativity. In the absence of a 
magnetic field and for a constant entropy, the energy release takes 
place only in an extended disk and on the surface of a cool star.

To accurately describe the external field of a rotating neutron star, we 
proposed (SS 98) to use exact solutions of the Einstein equations in a 
vacuum, for which the Ernst complex potential on the symmetry axis has a 
simple structure with arbitrary constants that have the meaning of 
multipole coefficients. The exact quadrupole solution accurately 
describes the external fields of rapidly rotating neutron stars, and, by 
contrast to asymptotic expansions in inverse powers of the radius, is 
meaningful up to the stellar surface. Note that expanding our solution 
in inverse powers of the radius in the equatorial plane yields a result 
that differs from the result of Shibata and Sasaki (1998) by terms of 
the order of $j^4$, because their assumption about the order of 
Geroch - Hansen's multipole coefficient $Q_4$ breaks down in our exact 
solution.

As the NS rest mass approaches the largest mass which is stable only in 
the presence of rotation, the quadrupole coefficient decreases, and the 
NS external gravitational field differs only slightly from the 
gravitational field of a rotating black hole (Kerr solution).

CONCLUSION

The method developed here and based on the static criterion for 
stability has allowed us to construct approximation formulas for the NS 
mass and its quadrupole coefficient as functions of its rotation 
parameter $j$ and rest mass $m.$ Using these functions and their partial 
derivatives, we determined all the remaining global NS characteristics: 
angular velocity, moment of inertia, equatorial radius, external 
gravitational field, parameters of the marginally stable orbit at $R < 
R_*$, and parameters of an equatorial Keplerian orbit at $R> R_*$ as 
functions of $j$ and $m$ or as functions of $j$ and $M.$ To this end, we 
used the thermodynamic ideas developed in SS 00. To describe the 
external field of a NS, we used an exact solution of the Einstein 
equations with an additional constant compared to the Kerr solution, 
which has the meaning of quadrupole coefficient. In the parameter plane, 
we found the curves that separate the states with a NS inside and 
outside the marginally stable orbit, as well as the neutral curves of 
stability loss according to the static criterion.

The inferred parameters enabled us to calculate; (i) the energy release 
in the disk and on the NS surface in the presence and absence of a 
marginally stable orbit, (ii) the tetrad velocity components and the particle angular velocity on the NS surface on which they fell from the 
marginally stable orbit, and (iii) the spinup (spindown) rate as a 
function of the NS rest mass (or gravitational mass) and rotation 
frequency. We provided the corresponding approximation formulas for 
several equations of state at  $M =1.4\, M_{\odot}$. 

In the Newtonian formulation (when the stellar matter is modeled by an 
incompressible fluid), the surface energy release and the spindown rate 
are given in the form of Taylor expansions in terms of the dimensionless 
rotation frequency for an arbitrary NS mass.

Using the property of the solution for the external gravitational field 
to belong to the class of solutions for two coaxially rotating black 
holes or Kerr disks, we hypothesize that the NS stability to the 
"quadrupole" oscillation mode is lost at the critical quadrupole moment 
 $b = 0.25(1 -j^2)$.

ACKNOWLEDGMENTS

We wish to thank N. Stergioulas for his numerical code , N.Inogamov and R. 
Wielebinski for a discussion. We express our gratitude to Yu. Astakhov for
the help of translating  this paper in English. 

REFERENCES

1. M. A. Alpar, A. F. Cheng, M. A. Ruderman, and J. Shaham, Nature ,
v.300, 728 (1982).

2. W. D. Arnett and R. L. Bowers, Astrophys. J., Suppl. Ser. v.33, 415 
(1977).

3. J. M. Bardeen, W. H. Press, and S. A. Teukolsky, Astrophys. J. v.178, 
347 (1972).

4. J. M. Bardeen and R. V. Wagoner, Astrophys. J. v.167, 359 (1971).

5. G. Biehle and R. D. Blanford, Astrophys. J. v.411, 302 (1993).

6. L. Bildsten, astro-ph/0001135.

7. G. S. Bisnovaty - Kogan and S. I. Blinnikov, Astron. Astrophys. v.31, 
391 (1974).

8. G. S. Bisnovaty - Kogan and B. V. Komberg, Astron. Zh. v.51, 373 
(1974) [Sov. Astron. 18, 217 (1974)].

9. S. Bonazzola, E. Gourgoulhon, M. Salgado, and J. A. Marck, Astron. 
Astrophys. v.278, 421 (1993).

10. S. Bonazzola and J. Schneider, Astrophys. J. v.191, 273 (1974).

11. L. Burderi, A. Possenti, M. Colpi,, Di Salvo T., D'Amico
N., astro-ph/9904331.

12. I. M. Butterworth and J. R. Ipser, Astrophys. J. v.204, 200 (1976).

13. G. Calamai, Astrophys. Space Sci. v.8, 53 (1970).

14. F. Camilo, D. R. Lorimer, P. Freire, Lyne A. G., Manchester
R. N. , Astrophys. J., v.535, 
975 (2000); astro-ph/9911234.

15. D. Chakrabarty and E. H. Morgan, Nature v.394, 346 (1998).

16. S. Chandrasekhar, Ellipsoidal Figures of Equilibrium (Dover, 
New-York, 1986).

17. G. B. Cook, S. L. Shapiro, and S. A. Teukolsky, Astrophys. J. v.424, 
823 (1995).

18. B. Datta, A. V. Thampan, and I. Bombaci, Astron. Astrophys. v.334, 
943D (1998); astro-ph/9801312.

19. N. A. Dmitriev and S. A. Kholin, Vopr. Kosmog. v.9, 254 (1963).

20. K. Ebisawa, K. Mitsuda, and T. Hanawa, Astrophys. J. v.367, 213 
(1991).

21. Y. Eriguchi and E. Mueller, Astron. Asrophys. v.146, 260 (1985).

22. Y. Eriguchi, I. Hachisu, and K. Nomoto, Mon. Not. R. Astron. Soc. v.266, 179 (1994).

23. F. J. Ernst, Phys. Rev. v.50, 4993 (1995).

24. L. S. Finn and S. Shapiro, Asrophys. J. v.359, 444 (1991).

25. G. Fodor, C. Hoenselaers, and Z. Perjes, J. Math. Phys. v.30, 2252 
(1989).

26. B. Friedman and V. R. Pandharipande, Nucl. Phys. A v.361, 502 
(1981).

27. J. F. Friedman, J. R. Ibser, and L. Parker, Astrophys. J. v.304, 115 
(1986).

28. M. Gilfanov, M. Revnivtsev, R. Sunyaev, and E. Churazov, Astron. 
Astrophys. v.339, 483 (1998).

29. I. Hachisu, Astrophys. J., Suppl. Ser. v.61, 479 (1986).

30. I. Hachisu, Y. Eriguchi, and D. Sugimoto, Prog. Theor. Phys. v.68, 
191 (1982).

31. R. O. Hansen, J. Math. Phys. v.15 (1), 46 (1974).

32. J. B. Hartle, Phys. Rep. v.46, 202 (1978).

33. J. B. Hartle and D. H. Sharp, Astrophys. J. v.147, 317 (1967).

34. J. B. Hartle , Astrophys. J. v.195, p.203 (1975).

35. J. B. Hartle and K. S. Thorne, Astrophys. J. v.158, 719 (1969).

36. C. Hoenselaers, Prog. Theor. Phys. v.72, 761 (1984).

37.  N.A.Inogamov and R.A.Sunyaev, Pis'ma Astron. Zh., v.25, p.323
(1999); astro-ph 9904333. 

38.  A. F. Illarionov and R. A. Sunyaev, Astron. Zh. v.51, 1162 (1974) 
[Sov. Astron. 18, 691 (1974)].

39.  W.Kley, Astron. Astrophys. v.247, p.95 (1991)

40.  W.Kluzhniak, Ph.D. Thesis, Stanford Univ. 1987.

41. W. Kluzniak and R. V. Wagoner, Astrophys. J. v.297, 548 (1985).

42. H. Komatsu, Y. Eriguchi, and I. Hachisu, Mon. Not. R. Astron. Soc. 
v.237, 355 (1989).

43. D. Kramer and G. Neugebauer, Phys. Lett. A v.75, 259 (1980).

44. W. Laarakkers and E. Poisson, gr-qc/9709033.

45. M. J. Lighthill, Mon. Not. R. Astron. Soc. v.110, 339 (1950).

46. L. Lindblom, Phys. Rev. D 58, 024008 (1998); gr-qc/9802072.

47. V. M. Lipunov and K. A. Postnov, Astrophys. Space Sci. v.106, 103 
(1984).

48. C. P. Lorenz, D. G. Ravenhall, and C. J. Pethick, Phys. Rev. Lett. 
v.70, 379 (1993).

49. V. S. Manko, E. W. Mielke, and J. D. Sanabria-Gomez, Phys. Rev. D 
v.61, 081501 (2000); gr-qc/0001081.

50. V. S. Manko, J. Martin, E. Ruiz, et al., Phys. Rev. D v.49, 5144 
(1994).

51. V. S. Manko and E. Ruiz, Class. Quantum. Grav. v.15, 2007 (1998).

52. D. Markovic and F. K. Lamb, Rossi2000: Astrophysics with the Rossi 
X-ray Timing Explorer (NASA's Goddard Space Flight Center, Greenbelt, 
2000), p. E61.

53. S. L. W. McMillan and P. Hut, Astrophys. J. v.467, 348 (1996).

54. M. C. Miller and F. K. Lamb, Astrophys. J. v.470, 1033 (1996).

55. M. C. Miller, F. K. Lamb, and G. B. Cook, Astrophys. J. v.509, 793 
(1998); astro-ph/9805007.

56. G. Neugebauer, J. Phys. v.13, L19 (1980).

57. I. D. Novikov and V. P. Frolov, Physics of Black Holes (Nauka, 
Moscow, 1986).

58. T. Nozawa, N. Stergioulas, E. Gourgoulhon, and Y. Eriguchi, Astron. 
Astrophys. v.132, 431N (1998); gr-qc/9804048.

59. J. P. Ostriker and J. W.-K. Mark, Astrophys. J. v.151, 1075 (1968).

60. V. R. Pandharipande, Nucl. Phys. A v.174, 641 (1971).

61.  R.Popham and R. Narayan, Astrophys. J., v.442, p.337 (1995).

62. P. Popham and R. Sunyaev, astro-ph/0004017.

63. J. E. Pringle and M. J. Rees, Astron. Astrophys. v.21, 1 (1972).

64. W. H. Ramsey, Mon. Not. R. Astron. Soc. v.110, 325 (1950).

65. D. G. Ravenhall and C. J. Pethick, Astrophys. J. v.424, 846 (1994).

66. R. Ruffini and J. A. Wheeler, Bull. Am. Phys. Soc. v.15 (11), 76 
(1970).

67. F. D. Ryan, Phys. Rev. D v.52, 5707 (1995).

68. F. D. Ryan, Phys. Rev. D v.55, 6081 (1997).

69. M. Salgado, S. Bonazzola, E. Gourgulhon, and P. Haensel, Astron. 
Astrophys. v.291, 155 (1994).

70. Z. F. Seidov, Astron. Zh. v.48, 443 (1971) [Sov. Astron. v.15, 347 
(1971)].

71. N. I. Shakura and R. A. Sunyaev, Adv. Space Res. v.8 (2-3), 135 
(1988).

72. S. L. Shapiro and S. A. Teukolsky, Black Holes, White Dwarfs, and 
Neutron Stars: the Physics of Compact Objects (Wiley, New York, 1983; 
Mir, Moscow, 1985).

73. M. Shibata and M. Sasaki, Phys. Rev. D v.58, 104011 (1998); 
gr-qc/9807046.

74. N. R. Sibgatullin, Oscillations and Waves in Strong Gravitational 
and Electromagnetic Fields (Nauka, Moscow, 1984; Springer-Verlag, 
Berlin, 1991).

75. N. R. Sibgatullin and R. A. Sunyaev, Pis'ma Astron. Zh. v.24, 894 
(1998) [Astron. Lett. 24, 774 (1998)]; astro-ph/9811028.

76. N. R. Sibgatullin and R. A. Sunyaev, Pis'ma Astron. Zh. (2000) (in 
press) [Astron. Lett. (2000) (in press)].

77. H. C. Spruit and E. S. Phinney, Nature v.393, 139 (1998)

78. N. Stergioulas, http://pauli.phys.uwm.edu/Code/rns; 
www.livingreviews.org/ Articles/Volume1/1998-8stergio.

79. N. Stergioulas and J. L. Friedman, Astrophys. J. v.444, 306 (1995).

80. T. Stromayer, W. Zhang, J. H. Swank, et al., Astrophys. J. Lett. 
v.498, 1358 (1998); astro-ph/03119.

81. R. A. Sunyaev and N. I. Shakura, Pis'ma Astron. Zh. v.12, 286 (1986) 
[Sov. Astron. Lett. v.12, 117 (1986)].

82. J. L. Tassoul, Theory of Rotating Stars (Princeton Univ. Press,
Princeton, 1978).

83. A. Thampan and B. Datta, Mon. Not. R. Astron. Soc. v.297, 570
(1998).

84. S. E. Thorsett and D. Chakrabarty, Astrophys. J. v.512, 288 (1999); 
astro-ph/9803260.

85. L. Titarchuk and V. Osherovich, astro-ph/0005375.

86. M. Van der Klis, astro-ph/0001167.

87. R. A. D. Wijnands and M. van der Klis, Astrophys. J. Lett. v.482, 
L65 (1997).

88. R. B. Wiringa, V. Fiks, and A. Fabroccini, Phys. Rev. C v.38, 1010 
(1988).

89. M. N. Zaripov, N. R. Sibgatullin, and A. Chamorro, Prikl. Mat. Mekh. 
v.59 (5), 750 (1995).

90. M. N. Zaripov, N. R. Sibgatullin, and A. Chamorro, Vestn. Mosk. 
Univ., Ser. 1: Math., Mech., No. 6, 61 (1994).

91. Ya. B. Zel'dovich, Vopr. Kosmog. v.9, 157 (1963).

92. Ya. B. Zel'dovich and I. D. Novikov, The Theory of Gravitation and 
Evolution of Stars (Nauka, Moscow, 1971).

 \end{document}